\documentclass{pasj00}
\draft

\newcommand{\asca}{{\it ASCA\/}}
\newcommand{\rosat}{{\it ROSAT\/}}
\newcommand{\chandra}{{\it Chandra\/}}

\begin{document}
\SetRunningHead{H. Katayama and K. Hayashida}{}
\Received{}
\Accepted{}

\title{X-ray Study of the Dark Matter Distribution in Clusters of
Galaxies with Chandra}

\author{Haruyoshi \textsc{Katayama}}%
\affil{Japan Aerospace Exploration Agency, 2-1-1, Sengen, Tsukuba, Ibaragi 305-8505}
\email{hkatayam@oasis.tksc.jaxa.jp}

\and

\author{Kiyoshi \textsc{Hayashida}}
\affil{Graduated School of Osaka University, 1-1, Machikaneyama, Toyonaka,
Osaka 560-0043}\email{hayasida@ess.sci.osaka-u.ac.jp}



\maketitle

\begin{abstract}

We study the total gravitating mass distribution in the central region
of 23 clusters of galaxies with \chandra.  Using a new deprojection
technique, we measure the temperature and gas density in the very
central region of the clusters as a function of radius without assuming
any particular models. Under the assumptions of hydrostatic equilibrium
and spherical symmetry, we obtain the deprojected mass profiles of
these clusters.

The mass profiles are nicely scalable with a characteristic radius
($r_{200}$) and mass ($M_{200}$) on the large scale of
$r>0.1r_{200}$. In contrast, the central ($r<0.1r_{200}$) mass profiles
have a large scatter even after the scaling.  The inner slope $\alpha$
of the total mass density profile ($\rho(r) \propto r^{-\alpha}$) is
derived from the slope of the integrated mass profile. The values of the
inner slope $\alpha$ at the radius of $0.02r_{200}$ ($\alpha_0$) span a
wide range from 0 to 1.2.  For 6 out of 20 clusters, $\alpha_0$ is lower
than unity at a 90 \% confidence level. CDM simulations predict that the
inner slope $\alpha$ is in the range $1<\alpha<2$, which is inconsistent
with our results.  We also found that the gas fraction near the center
of a cluster has a negative correlation with $\alpha_0$. Our result
suggests that the gas-rich clusters in the central region tend to have a
flat core.

\end{abstract}

\section{Introduction}

The Cold Dark Matter (CDM) model has become the standard paradigm for
explaining observations of the large-scale structure of the
universe. In the CDM model, dark matter consists of non-baryonic,
collisionless, cold particle. The properties of dark matter density
profiles in the CDM model have been investigated extensively through
numerous N-body simulations. \citet{navarro:1997} (hereafter NFW)
claimed that the dark matter density profiles in the CDM model are
reasonably approximated by a universal form with singular behavior in
its central region. Several N-body simulations predict that the density
of dark matter increases as a power law $\rho(r) \propto r^{-\alpha}$,
with $\alpha$ in the range of 1 to 2, in the central region
(e.g., $\alpha = 1$ by NFW; $\alpha = 1.5$ by
\citet{moore:1998}). Measurements of the inner slope $\alpha$ of dark
matter density profiles offer a powerful test of the CDM model.

The observational efforts in this respect have been in the form of
dynamical studies of low surface brightness and dwarf galaxies. The
observations obtained in those studies suggest the presence of a
relatively flat core: $0<\alpha<1$ (e.g.,
\citet{firmani:2001}). Gravitational lensing has made some observational
constraints available at the scale of galaxy clusters. For instance,
\citet{sand:2002} showed that steep inner slopes ($\alpha>1$) are ruled
out at better than 99 \%, for the lensing cluster MS2137-23. Although
gravitational lensing studies provide a unique and important probe of
dark matter profiles, they generally can be applied only to a limited
sample of clusters that satisfy a specific lensing condition.  X-ray
observations of the density and temperature of a hot intracluster medium
(ICM), on the other hand, probe the mass of a cluster of galaxies under
the assumption of hydrostatic equilibrium. This could be a powerful tool
to investigate dark matter profiles in the central regions of
clusters. However, for previous X-ray satellites, such as \rosat\ and
\asca, the detailed study of ICM temperature and density profiles at
small scales has been difficult because of limitations on the
performance of imaging or spectroscopic instruments. The high spatial
resolution imaging spectroscopy of \chandra\ enables the measurement of
mass profiles in the very central regions of clusters of
galaxies. Several groups have obtained X-ray constraints on the dark
matter profiles of some clusters. These results are apparently
consistent with the CDM model (e.g., \citet{david:2001}, and
\citet{arabadjis:2002}). However, \citet{ettori:2002} showed that the
mass profile of A1795 flattens within 100 kpc. Systematic studies are
thus required for a large sample of clusters. In this paper, we
systematically study the mass profiles of 23 clusters of galaxies.


 We assume $\Omega_m = 1$, $\Omega_\lambda = 0$, and $H_0 =
50$ km s$^{-1}$ Mpc$^{-1}$ throughout this paper. Unless otherwise
noted, all errors are 1$\sigma$ (68.3 \%) confidence intervals.

\section{Sample and Analysis}
\label{section:sample_ana}

We selected our sample from \chandra\ archival data of galaxy
clusters. To obtain spatially resolved spectra, we restricted the
observations to those in which ACIS were employed without gratings. By
the end of September 2002, the archive contained observation data for
about 150 clusters ($\sim$ 200 pointings) that met this criterion.

We applied the following criteria for further selection of the data in
order to meet our main concern, investigation of the central mass
profiles of galaxy clusters. First, clusters must be bright enough to
provide spatially resolved spectra with good statistics. For this we
referred to the catalog of \citet{reiprich:2002}, which consists of 106
bright clusters compiled from several catalogs based on the \rosat\
All-Sky X-ray Survey (\citet{voges:1999}). The minimum X-ray flux among
the Reiprich samples is $0.234 \times 10^{-11}$ ergs s$^{-1}$ cm$^{-2}$
(0.1--2.4 keV). This is bright enough for our analysis under typical
observational conditions.  Among the 106 clusters in Reiprich sample, 43
clusters are included in the \chandra\ data archive. Secondly, clusters
should be spherically symmetric as our deprojection analysis assumes the
spherical symmetry. We thus excluded merging clusters like A754
(\citet{henriksen:1996}).  Although this second criterion is somewhat
ambiguous, we will examine how this spherical symmetry assumption
affects the final result in Section
\ref{section:obs_effect}. The last criterion is that the X-ray
emission from the outer region of a cluster must be covered by the
detectors used in the observations. This is because the deprojection
analysis depends on accurate measurement of temperatures and densities
of the outer regions of clusters. Data for 20 clusters met all of these
criteria. We also employed three distant clusters, A1835, A963, and
ZW3146, that are bright and spherically symmetric but that are not
included in \citet{reiprich:2002}.

 The observation log and the properties of each cluster are summarized
in Table \ref{tbl:obs_log} and Table \ref{tbl:sample_properties},
respectively. The redshifts of the 23 sample clusters range from 0.0110
to 0.2906, with a median of 0.0852.

Data reduction and analysis were performed with the \chandra\
Interactive Analysis of Observations package, CIAO-2.2, with calibration
database CALDB-2.12, as provided by the \chandra\ X-ray Center (CXC). We
started the reduction from the standard level 2 event files archived at
CXC, which are the products of the pipeline processing. We adopted the
standard reduction scheme by following the CIAO
threads\footnote{http://asc.harvard.edu/ciao/threads}.  

To remove the flare events, we performed lightcurve screening using the
CIAO task {\it lc\_clean}.  We made a background lightcurve, a time
history of the event rate taken from a source-free region on the
detector, with a time bin size of 259.28 s using the CIAO task {\it
lightcurve}. In order to exclude flare events, we discarded the data
taken at the time the count rate deviates from the mean by $\pm3\sigma$,
where the mean value is defined during the quiescent period.  Point
source detection was performed with the CIAO wavelet source detection
routine {\it wavdetect} with a significance parameter $10^{-6}$. We made
a 0.3--10 keV image binned by using a bin size of about $2\times2''$
($4\times4$ pixels). The area around the detected point sources was
excluded in the following analysis.  In order to estimate the background
level to be subtracted from the X-ray spectra and images, we applied the
blank-sky data compiled by Markevitch (2001)\footnote{http://cx
c.harvard.edu/contrib/maxim/bg/index.html} as background data. These
background data are event files made with the same lightcurve screening
process of the cluster data.

The spectra were extracted in the concentric annuli centered on the
X-ray peak with different widths to ensure similar statistics in the
background-subtracted spectra.  The X-ray peak was determined with the
X-ray images from which point sources were removed. We examine what the
appropriate setup is for the width of the annuli, or equivalently, the
statistics of each spectrum, using a simulation. 
From this simulation, we found that the lower photon counts or the
more annulus result in the larger systematic errors in the temperature
profile as shown in \citet{arabadjis:2002}. 
In order to suppress the systematic errors in the temperature profile,
we restricted the photon count per each annulus and the number of
annuli, as follows: (1) The photon count per each annulus must be at least
$1\times10^4$, and (2) The number of annuli $N$ must be $5 \leq N \leq
10$. The radius of the outermost annulus was determined to cover the
4-$\sigma$ background level of the \rosat\ PSPC image. 
The average of the outermost radius is about 720 kpc. The
background spectra were extracted from the background data with the same
regions on the detector.  Redistribution Matrix Files (RMF) and
Auxiliary Response Files (ARF) were made using the CIAO tasks of {\it
mkrmf\/} and {\it mkwarf\/}. These tasks make a weighted RMF and a
weighted ARF for the spectral analysis based on a $32\times32$ pixel
grid of calibration files. This is because the RMF and ARF vary with
detector location. To compensate for the degradation in low-energy
efficiency, we used the tool {\it corrarf\/} provided by CXC. The {\it
corrarf\/} corrects the ARF according to the observation date.

\subsection{Deprojection Analysis}
\label{subsection:deproject}

To determine the deprojected temperature and gas density profiles, we
applied a new deprojection technique developed by
\citet{arabadjis:2002}. We here briefly summarize this technique (see
also \citet{arabadjis:2002} and \citet{katayama:2003}). 

\begin{figure}[h]
 \begin{center}
  \FigureFile(140mm,100mm){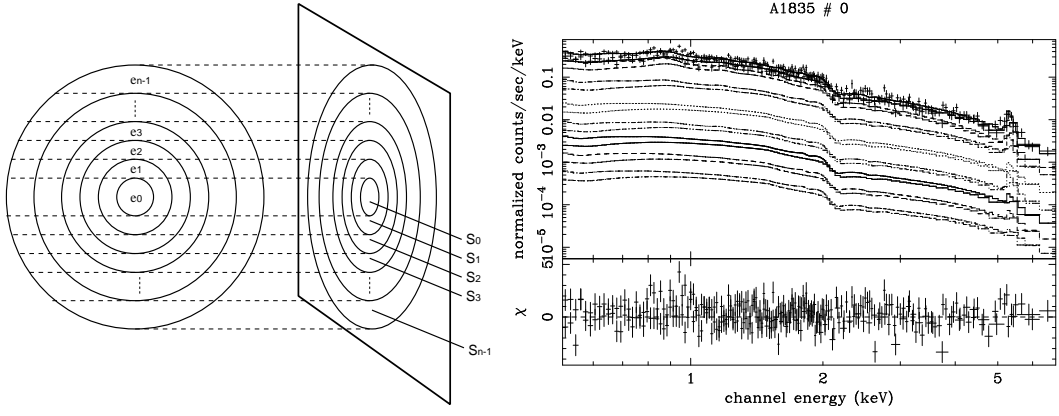} 
 \end{center}
\caption{(Left) Schematic view of the deprojection analysis. The left
spheres represent X-ray emissions from $N$ spherical shells, and the right circles represent observed projected luminosities which are the
integrations of emissivities along the line of sight. (Right) An example
of the spectrum fitting. In this example (A1835), we fitted the $N = 7$
spectra for two data sets simultaneously. Observation IDs of each data
set are 495 and 496.}
  \label{fig:deproject}
\end{figure}

A Schematic view of the deprojection analysis is shown in
Figure \ref{fig:deproject} (Left). In this example, we extract spectra from $N$
concentric annular regions.  The projected luminosity $S_j$ in a given
energy band on the $j$th annulus is expressed by the integration of
emissivities along the line of sight.  The relationship between $S_j$
and the volume emissivity $e_i$ of the $i$th spherical shell is
expressed as
\begin{equation}
 S_j = \sum_{i=j}^{N} V_{i j} e_i.
\label{eqn:deproject0}
\end{equation}
where $V_{i j}$ is the volume of the $i$th spherical shell intersected by a
cylindrical shell whose radius equals the $j$th projected annulus. 
Note that we have to make sure, or make the assumption that, X-ray emission
is negligible outside of the outermost annulus. 
$V_{i j}$ is geometrically calculated as
\begin{eqnarray}
 \nonumber V_{i j} & = & \frac{4}{3}\pi [(r_{i+1}^{2}-b_{j}^{2})^{3/2}-(r_{i+1}^{2}-b_{j+1}^{2})^{3/2}-(r_{i}^{2}-b_{j}^{2})^{3/2}+(r_{i}^{2}-b_{j+1}^{2})^{3/2}] ~~ (i \geq j) \\
& = & 0 ~~ (i < j),
\end{eqnarray}
where $r_{i}$ and $r_{i+1}$ are the inner and outer radii of the $i$th
spherical shell, and $b_{j}$ and $b_{j+1}$ are the inner and outer radii
of the $j$th annulus, which equal $r_{j}$ and $r_{j+1}$, respectively.
Since Equation \ref{eqn:deproject0} can also be written for all annuli,
these are written as
\begin{equation}
 {\bf S} = {\bf V} \cdot {\bf e}.
\label{eqn:deproject_matrix}
\end{equation}
Since ${\bf V}$ is a triangle matrix, we can obtain the ${\bf e}$ 
by solving the inverse matrix ${\bf V}^{-1}$. 

In most previous deprojection analyses, only X-ray spatial information
was utilized and an additional assumption on the temperature profile
$kT(r)$ or the potential profile $\phi(r)$ was necessary. Even when both
types of information are available, some authors assume the potential
profile $\phi(r)$, or equivalently, the gravitational mass density
profile $\rho(r)$, beforehand. However, the method by
\citet{arabadjis:2002} does not make such assumptions. We first make a
trial model for volume emissivity $e_i$ at each spherical radius, which
is a function of gas temperature $kT_i$, gas density $n_{g,i}$ and gas
abundance $Z_i$ when we employ an X-ray emissivity model of thin thermal
plasma.  We adopted the MEKAL
\citep{mewe:1985,mewe:1986,kaastra:1993,liedahl:1995} model in the XSPEC
data analysis package for our X-ray emissivity model, in which
normalization $K_i$ is used instead of gas density $n_{g,i}$. Therefore,
the number of free parameters to be determined is 3$\times N$ except for
an additional free parameter for the interstellar absorption $N_{\rm
H}$.  We can examine how this trial model fits the set of spectra by
$\chi^2$ value, and we can improve the fit by changing the parameter
values $kT_i$, $n_{g,i}$, and $N_{\rm H}$. This procedure is done with
the XSPEC data analysis package as a simultaneous spectral fitting of
$N$ spectra. For some clusters, the interstellar absorption was poorly
constrained. In such case, we fixed the absorption column to the
Galactic value.  We show the sample of the fitting result in Figure
\ref{fig:deproject} (Right).

\section{Temperature and Gas Density Profiles}
\label{section:temp_ne_p}

\begin{figure}[h]
 \begin{center}
  \FigureFile(70mm,50mm){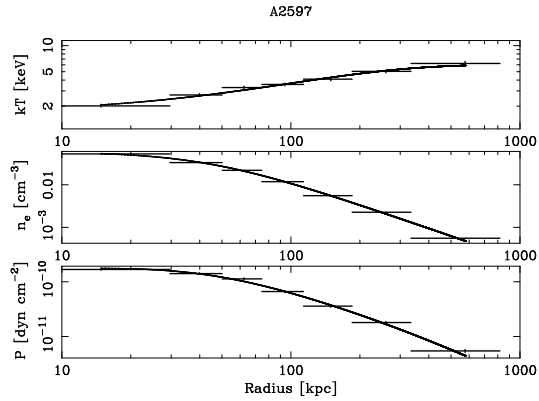}
 \end{center}
  \caption{The temperature, electron density, and pressure profiles of
  A2597. The solid lines represent best-fit models with analytical functions (Equation \ref{eqn:temp_fit_exp_model} and Equation \ref{eqn:nfw_gas}). The pressure is simply derived from the temperature and gas density with the equation of state of the ideal gas: $P = n_{e}kT$.}
  \label{fig:temp_ne_presure_a2597}
\end{figure}

We determined the temperature and gas (electron) density profiles of all
sample clusters. Figure \ref{fig:temp_ne_presure_a2597} shows the
temperature, gas density, pressure profiles of A2597. The pressure
profiles are simply derived from the temperature and gas density with
the equation of the state of ideal gas: $P = n_{e}kT$.  We attempted to
model the temperature and density profiles with analytic functions. Note
that the total mass profile can be calculated without employing such
models, as shown in the next section.  However, we investigated for
another way to obtain the total mass profile, to which end such models
are employed.  We fitted the temperature profile with exponential +
constant model given by
 \begin{equation}
 T(r) = T_0 + T_1 {\rm exp} (-r/r_T).
 \label{eqn:temp_fit_exp_model}
 \end{equation} 
The fitting results are summarized in Table \ref{table:fit_temperature_exp}.
Three parameters in the exponential + constant model were determined by
the $\chi^2$ fitting, but their error estimation was not trivial when
the fitting was unacceptable.  
These large reduced $\chi^2$ values
are likely to be due to local fluctuations in the intrinsic temperature
profiles or to unknown systematic errors in our analysis procedure. Note
that we integrate the above model functions within one radius bin to obtain
each model point. Thus, a coarse sampling is not the cause of the
large $\chi^2$ values.  
In order to estimate conservative errors
for the parameters, we assigned a systematic error to each data point in
the temperature profile.  The systematic error of the temperature is
assumed to be the constant fraction of the measured temperature for all
the data points, where the fraction is determined so as to obtain the
reduced $\chi^2$ of unity in each temperature profile.  Note that the
total error is calculated to be the square root of the quadratic sum of
the statistical and systematic errors. For some clusters, the original
(i.e., before assignment of the systematic error) reduced $\chi^2$ is
small enough to accept the fit. We assigned the systematic error only
for the clusters for which the original fitting was rejected by the
$\chi^2$ test with a significance level less than 1 \%.
We also fitted the gas density profile with NFW gas density model given by
\begin{equation}
n_{g}(r) = n_{e0}~{\rm exp}[-B(1-\frac{{\rm ln}(1+(r/r_{\rm s}))}{(r/r_{\rm s})})]
\label{eqn:nfw_gas}
\end{equation}
\citep{makino:1998}.
The fitting results are summarized in Table
\ref{table:fit_gas_density_nfw}. As was the case for the temperature
profiles, we adopted the systematic error for all the clusters.

\section{Mass Profiles}
\label{section:int_mass}

Under the assumptions of hydrostatic equilibrium and spherical symmetry,
we can obtain the total gravitating mass profile as a function
of radius using
 \begin{eqnarray}
\nonumber \frac{dP_g}{dr} & = & - \mu n_g m_p \frac{d\phi}{dr} \\
                          & = & - \mu n_g m_p \frac{GM(<r)}{r^2}.
\label{eqn:hydrostatequib2}
\end{eqnarray}

We derived this mass profile by two different methods in order to check
its consistency.  The first method employs the temperature and density
profile models obtained in Section \ref{section:temp_ne_p}.  Substituting
Equation \ref{eqn:temp_fit_exp_model} and Equation \ref{eqn:nfw_gas} for
Equation \ref{eqn:hydrostatequib2}, we obtain the mass profile in an analytic
form.  The second method does not employ the temperature and density
profile models. Instead, the mass profile is derived by
approximating Equation \ref{eqn:hydrostatequib2} as simple differences:
\begin{equation}
 M(<r) \sim -\frac{1}{{\mu}n_g(r)m_p}\frac{r^2}{G}\frac{{\Delta}P(r)}{{\Delta}r}
.
\label{eqn:mass_diff}
\end{equation}
We calculated ${{\Delta}P(r)}/{{\Delta}r}$ as
\begin{equation}
 \frac{{\Delta}P(r)}{{\Delta}r} = \frac{P_{i+1}-P_{i}}{r_{i+1}-r_i}
\label{eqn:pressure_diff}
\end{equation}
where $P_{i}$ and $r_{i}$ are the pressure and radius of the $i$th
shell. 
The radius $r$ and gas density $n_g(r)$ are given by
$r=(r_{i+1}+r_{i})/2$ and $n_g(r)=(n_{e i+1}+n_{e i})/2$,
respectively.

\begin{figure}[h]
 \begin{center}
  \FigureFile(70mm,50mm){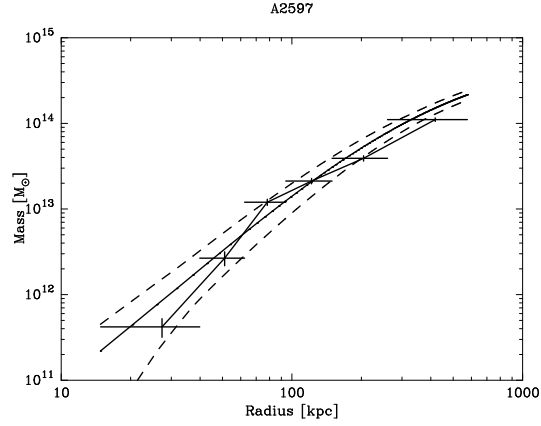}
 \end{center}
  \caption{The mass profiles of A2597.  The solid curves represent the mass profile derived from the best-fit parameters of Equation \ref{eqn:temp_fit_exp_model} and Equation \ref{eqn:nfw_gas}. The dashed lines are a confidence level of 68 \%. The discrete data with error bars represent the mass profile calculated using approximate expression given by Equation \ref{eqn:mass_diff}.  } 
  \label{fig:mass_profiles_a2597}
\end{figure}

Results of these two methods are compared in Figure
\ref{fig:mass_profiles_a2597} for A2597. The plots also show the
1$\sigma$ confidence levels for the analytic mass profile, which is
derived by considering the errors of the parameters describing the
temperature and density profile models.  The mass profiles derived by
the two different methods were consistent in most of the cases.  In the
case that the pressure of the outer shell is larger than that of the
inner shell ($P_{i+1}>P_{i}$), the mass profile shows the negative value
at that point : $M<0$. This is likely caused by the local temperature
fluctuation intrinsic at some radius of the clusters or by systematic
errors in our analysis. We excluded these unphysical points in the
analysis, though such points are only seen in NGC5044 and Centaurs. Of
the two derivation methods, the first one using the modeled temperature
and density profiles is easy to handle, but it involves sacrificing one
important point of the deprojection analysis; i.e., no reliance on any
particular profile models. On the other hand, the second method is more
straightforward, but it suffers large error owing to local fluctuations
in the temperature and density profiles.

\subsection{Scaling of Mass Profiles}
\label{section:scale_mass}

\begin{figure}
 \begin{center}
  \FigureFile(70mm,50mm){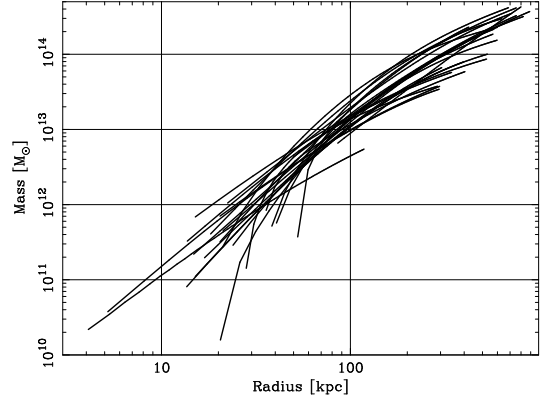}
 \end{center}
  \caption{Mass profiles of 23 sample clusters obtained from the temperature and density profile models. }
\label{fig:unscaled_mass_profiles}
\end{figure}

\begin{figure}
 \begin{center}
  \FigureFile(70mm,50mm){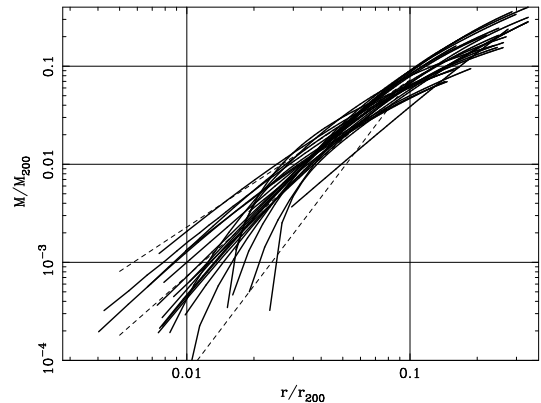} 
 \end{center}
 \caption{Mass profiles scaled by $M_{200}$ and $r_{200}$. For the
 calculation of $r_{200}$ and $M_{200}$, we used the relation obtained
 from the numerical simulation by \citet{evrard:1996}. The dashed lines
 represent $M{\propto}r^{1.5}$ ($\alpha=1.5$), $M{\propto}r^2$
 ($\alpha=1$) and $M{\propto}r^3$ ($\alpha=0$), respectively. }
\label{fig:scaled_mass_profiles}
\end{figure}

In Figure \ref{fig:unscaled_mass_profiles}, we show the analytic mass
profiles of 23 sample clusters in one plot, illustrating the scatters
among them.  CDM simulations predict that the density profiles of dark
matter are universal in form across a wide range of mass scales
\citep{navarro:1995,navarro:1996}.  We scaled our analytic mass profiles
with $r_{200}$ and $M_{200}$, where $r_{200}$ is the radius within which
the mean halo density is 200 times the critical density of the universe,
and $M_{200}$ is the total mass enclosed within $r_{200}$. As shown by
\citet{navarro:1995}, and \citet{navarro:1996}, clusters of different
mass are expected to show similar structures when scaled to such a
characteristic radius and mass.  For the calculation of $r_{200}$, we
used the relation obtained from the numerical simulation by
\citet{evrard:1996}:
\begin{equation}
 r_{200} = 3.690~(T/10 {\rm keV})^{0.5} ~(1+z)^{-1.5} ~~[{\rm Mpc}],
 \label{eqn:r200}
\end{equation}
where $T$ is the spatially averaged temperature, and $z$ is the redshift. 
$M_{200}$ is calculated by
\begin{equation}
 M_{200} = \frac{4}{3} \pi (200\rho_{\rm crit}(z)) ~r_{200}^3.
\label{eqn:m200}
\end{equation}
where $\rho_{\rm crit}(z) = 3H(z)^2/8{\pi}G$ is the critical density of the
universe at a redshift $z$. We show the scaled mass profiles in Figure
\ref{fig:scaled_mass_profiles}.  On a large scale ($r>0.1r_{200}$), the
scaled mass profiles agree with each other better than did the original
mass profiles, except in the case of one deviant profile of A401. This
findings suggest that the mass profiles have a similar form on a large
scale; in other words, the scaling with $r_{200}$ and $M_{200}$ is
effective at least on this scale.  The standard deviation of the mass
profiles is 41 \% at 200 kpc for the original mass profiles, and that
for the scaled profiles is 21 \% at $0.1r_{200}$, which corresponds to
about 160--300 kpc.  In contrast, the standard deviations on the small
scale ($r<0.1r_{200}$) are not significantly different: 55 \% at 20 kpc
for the original mass profiles, and 60 \% at $0.01r_{200}$ for the
scaled mass profiles.  

When the density profile of dark matter is described with the power-law
expression $\rho(r) = \rho_0 (r/r_0)^{-\alpha}$, the mass integrated over
the volume is described by
\begin{equation}
 M(<r) = \int_{0}^{r} 4\pi\rho(r^{\prime})r^{\prime 2} dr^{\prime} = \frac{4\pi\rho_0r_0^3}{3-\alpha}(\frac{r}{r_0})^{3-\alpha}.
\label{eqn:mass_pl}
\end{equation}
Therefore, the smaller the value of $\alpha$ is ($\alpha
\rightarrow 0$), the steeper the mass profile is. We
overlaid the $M \propto r^{1.5}$ ($\alpha = 1.5$), $M \propto r^2$ ($\alpha
= 1$) and $M \propto r^3$ ($\alpha = 0$) lines on the scaled mass
profiles in Figure \ref{fig:scaled_mass_profiles}. It was found that the slope
$\alpha$ was in the range of 0 to 1.5, and it was flatter (smaller) on the
small scale.  The slope $\alpha$ at the cluster center is quantitatively
examined in Section \ref{section:central_mass_slope}

\subsection{Inner Slope of Dark Matter Distribution}
\label{section:central_mass_slope}

The shape of the dark matter distribution near the center of a cluster
is sensitive to the theoretical models adopted.  In this section, we
focus on the observed shape of the total mass distribution in terms of
the slope of the density profile at the inner part of a cluster. The
inner slope of the density profile is obtained by fitting the total mass
profile we obtained with a model mass profile calculated from an assumed
density profile.  Although we employed the King and NFW profiles as
density profile models in the previous section, a more generalized form
is used in this section. That form was
\begin{equation}
 \rho(r) = \frac{\rho_0}{(r/r_{\rm s})^{\alpha}(1+(r/r_{\rm s}))^{(3-\alpha)}},
\label{eqn:general_dens_profile}
\end{equation}
where $\rho_0$ is the central density, $r_{\rm s}$ is its scale radius,
and $\alpha$ is the asymptotic slope of the profile at small radii. This
form of density profile requires numerical integration to derive the
integrated mass profile.  The asymptotic slope $\alpha$ in Equation
\ref{eqn:general_dens_profile} can be used as the inner slope. However,
it was found that the asymptotic slope $\alpha$ and the scale radius
$r_{\rm s}$ are coupled strongly, and therefore difficult to determine
independently. Thus, we focus on the slope of the density profile at a
finite radius and used it as the inner slope. The slope at a radius
$r_0$, $\alpha_0(r_0)$ is given by
\begin{equation}
 \alpha_0(r_0) \equiv - \frac{d{\rm ln}\rho(r)}{d{\rm ln}r}\mid_{r=r_0}.
 \label{eqn:def_alpha0}
\end{equation}
Using Equation \ref{eqn:general_dens_profile}, we get 
\begin{equation}
 \alpha_0(r_0) = \frac{(\alpha+({r_0}/{r_{\rm s}}))}{(1+({r_0}/{r_{\rm s}}))}.
 \label{eqn:alpha0}
\end{equation}
We employ $\alpha_0(r_0)$ instead of $\alpha$, in addition to $\rho_0$
and $r_{\rm s}$, as a free parameter of the fitting.  We fixed $r_0$ to
$0.02r_{200}$, which corresponds to about 40 kpc. The choice of $r_0$ is
not trivial, but we fix this value so that the radius is appropriate for
a comparison of the theoretical models and is covered by observed data
points in the mass profile.  Nevertheless, A401, A644, and A963, due to
a lack of data points within $0.02r_{200}$, were discarded from the
following analysis. Since the mass profiles of some clusters are nearly
power-law, we could not determine $r_{\rm s}$. In such case, we fixed
$r_{\rm s}$ to 1 Mpc, which is the radius that there is no data for all
clusters.

\begin{figure}
 \begin{center}
 \FigureFile(70mm,50mm){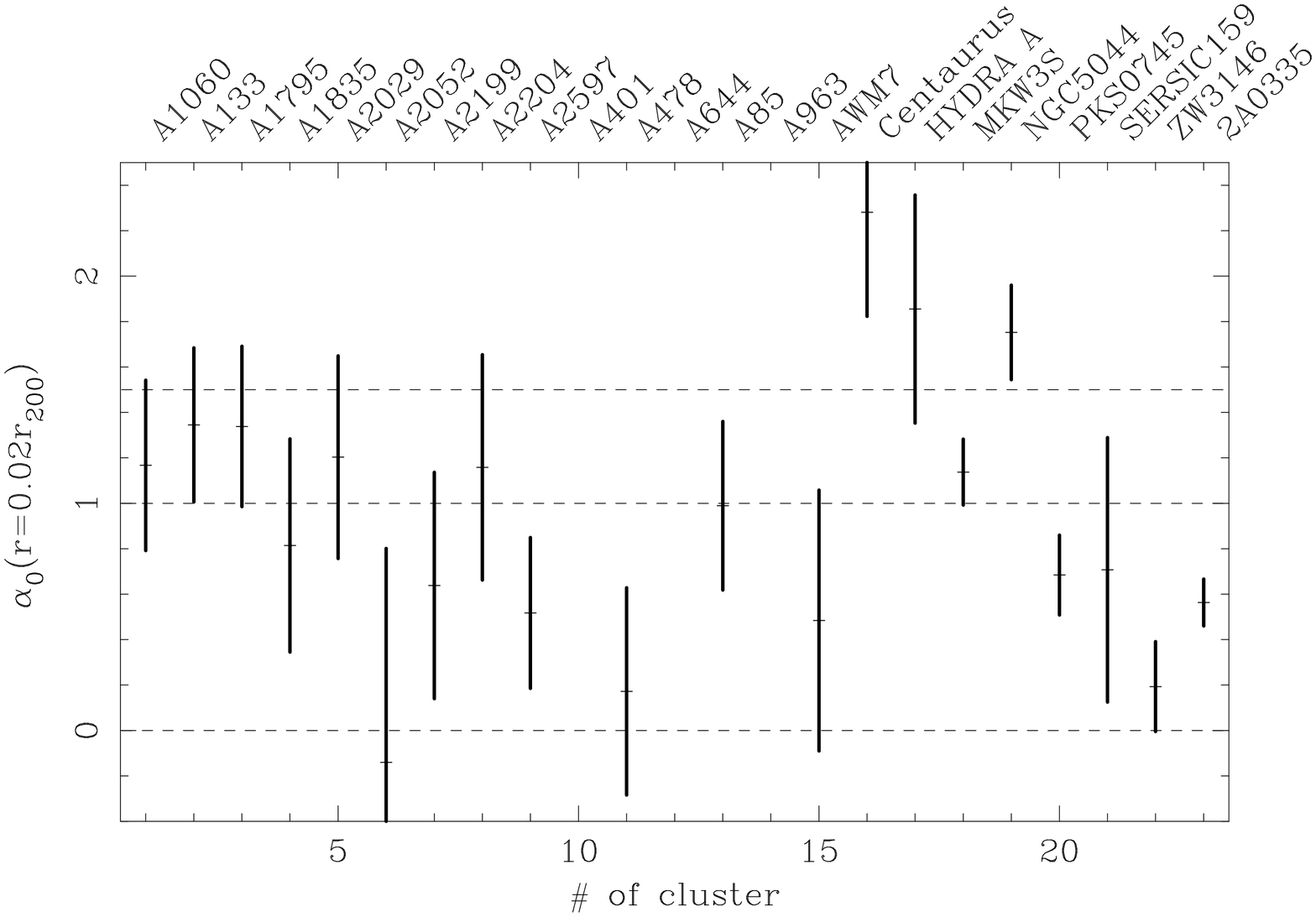}
 \end{center}
 \caption{Values of
 the Inner slope $\alpha_0$ at the radius of 0.02$r_{200}$ for 20
 clusters in Table \ref{tbl:alpha_002r200}. A401, A644, and A963 were removed
 due to a lack of data points within $0.02r_{200}$. The
 horizontal dashed lines represent $\alpha=1.5$ (Moore), $\alpha=1.0$
 (NFW), and $\alpha=0.0$ (King). Error bars are shown at 90 \%
 confidence level.}  
\label{fig:alpha_002r200}
\end{figure}

\begin{figure}
 \begin{center}
  \FigureFile(170mm,220mm){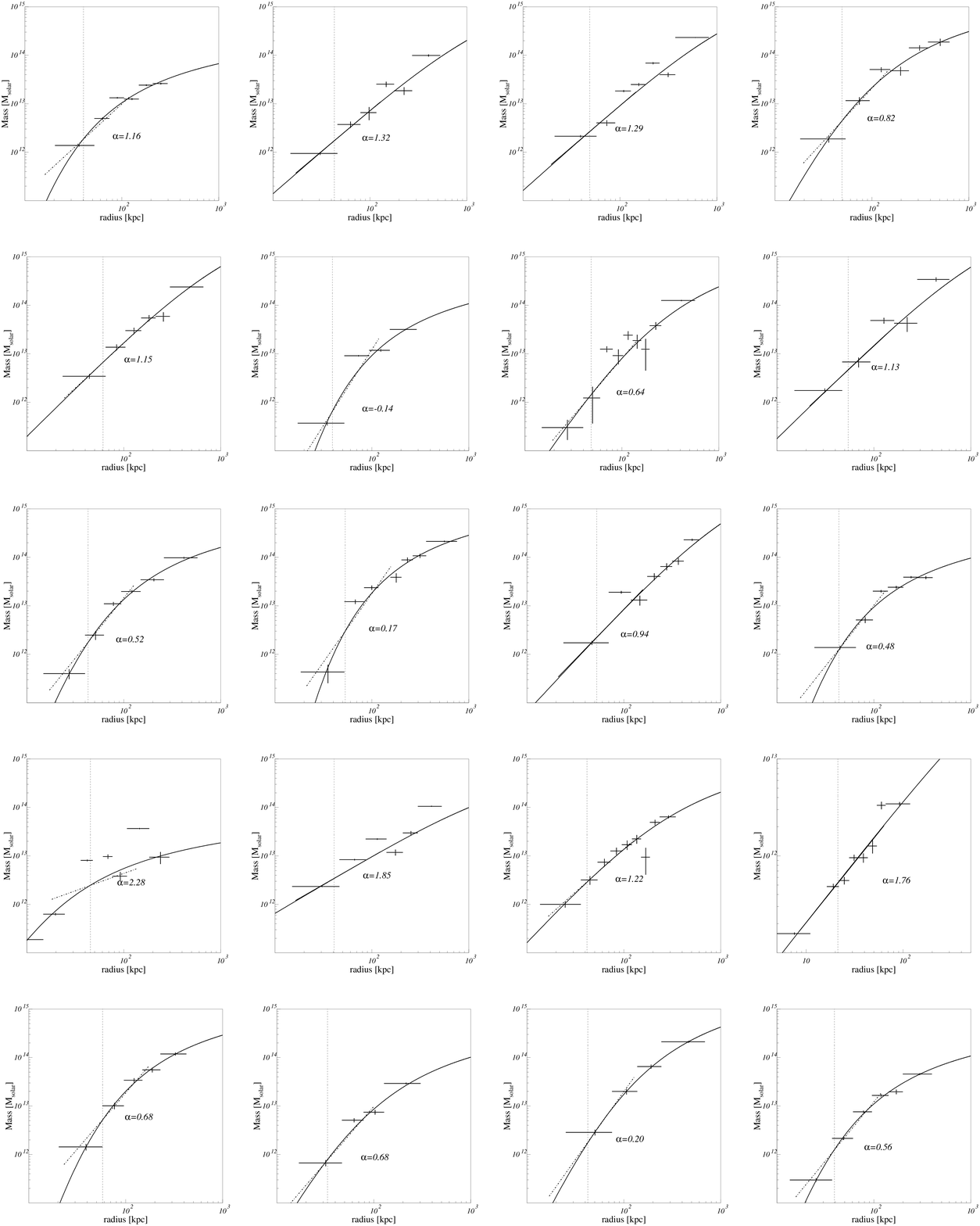} 
 \end{center}
 \caption{The best-fit mass models using the general form of the density
 profile. The horizontal dashed lines represent the radius of $r_0 =
 0.02r_{200}$. The dash-dotted lines represent the best-fit slope ($M
 \propto r^{3-\alpha_0}$ at the radius $r_0$. A401, A644, and A963 are
 removed because of the lack of the data within $0.02r_{200}$.}
 \label{fig:mass_fit_alpha0}
\end{figure}

The results of the fitting are summarized in Table \ref{tbl:alpha_002r200},
and the best-fit values and errors of the inner slope $\alpha_0(r_0)$
are plotted in Figure \ref{fig:alpha_002r200}.  We also show the total mass
profiles with the best-fit models for the general form of the density profile
in Figure \ref{fig:mass_fit_alpha0}. The inner slope $\alpha_0$ spans a
wide range with $0 \leq \alpha_0 \leq 2.3$. We found that the 90\% upper
bound of $\alpha$ was lower than unity for 6/20 ($\sim41$ \%) clusters
(A2052, A2597, A478, PKS0745-191, ZW3146, and 2A0335+096), suggesting
that the dark matter distribution in a significant fraction of clusters
was flatter than that in CDM halo models such as the NFW profile or the
Moore profile.

\section{Examination of the Systematic Effects} 
\label{section:obs_effect}

We have demonstrated that the inner slope $\alpha_0$ shows a large
scatter and is less than unity for 30\% of the clusters in our
sample. Before discussing these results in greater detail, we would like
to address their validity and the systematic effects that may affect the
measured inner slope from various points of view.

\subsection{Center Position}
\label{section:center}

We can define three types of positions as the center of a cluster: (1)
the X-ray emission peak of cluster hot gas; (2) the X-ray centroid; and
(3) the position of the Brightest Cluster Galaxy (BCG). We chose the
X-ray emission peak as the center of the annuli to extract
spectra. Since these three positions are slightly different, the
selection of the center position may affect our results.

\begin{figure}
 \begin{center}
  \FigureFile(70mm,50mm){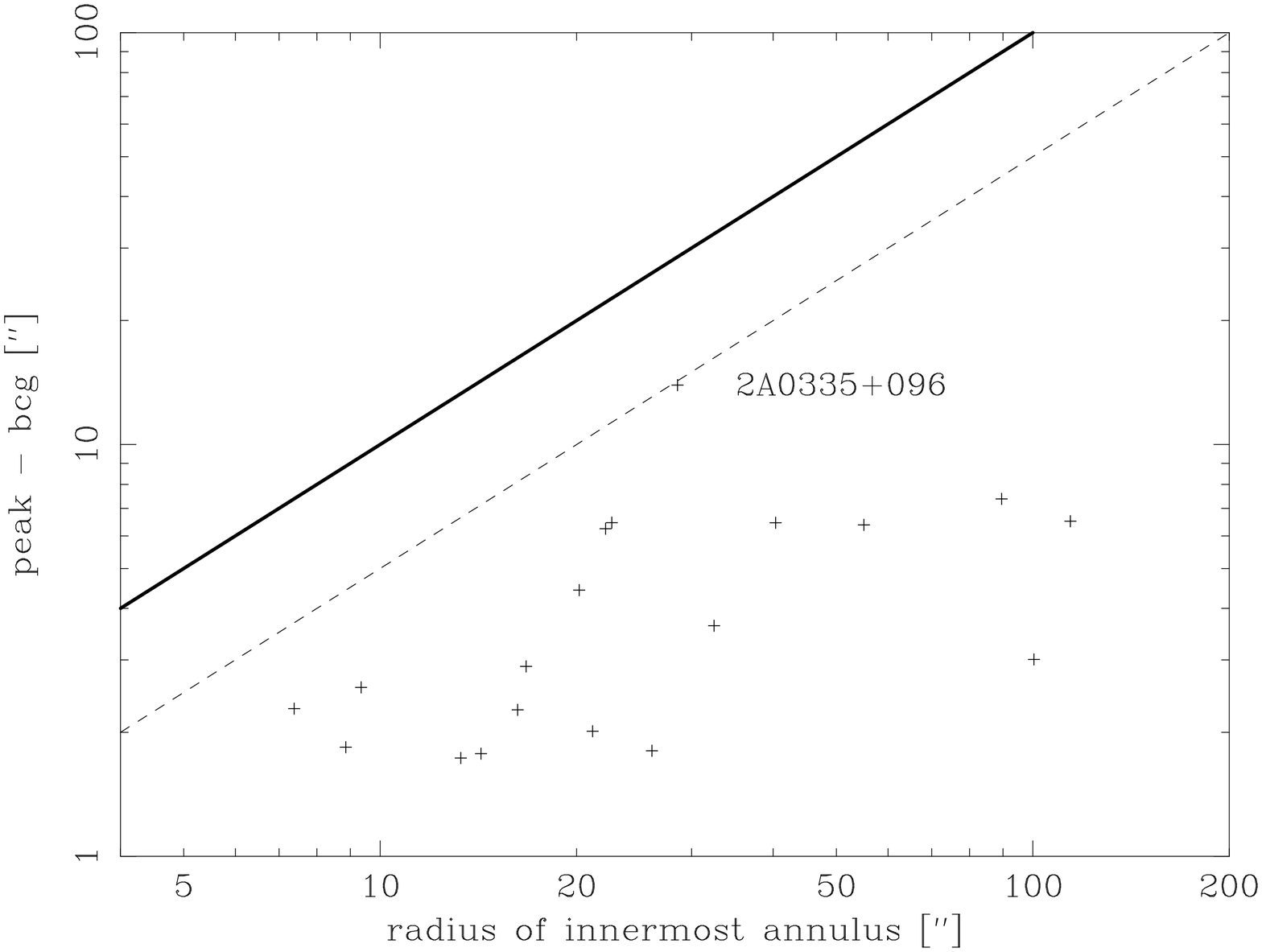}
 \end{center}
  \caption{Radius of the innermost annulus
  vs. offset between the X-ray peak and the BCG. The dashed line
  represents 50 \% of the radius of the innermost annulus.} 
  \label{fig:offset_bcg}
\end{figure}

We first examined the position of the BCG in our sample clusters. The
BCG is defined as the brightest galaxies among the member galaxies of a
cluster. Since the BCGs are usually located at the center of a cluster
and have velocities very near the mean velocity of galaxies in the
cluster, they are considered to sit at the bottom of the cluster
gravitational potential well. The positions of BCGs were taken from the
NASA/IPAC Extragalactic Database
(NED)\footnote{http://nedwww.ipac.caltech.edu}. We calculated the offset
between the X-ray peak and the position of the BCG. When the offset was
larger than or comparable to the radius of the innermost annulus, the
temperature or density profiles will be affected by the selection of the
center. In Figure \ref{fig:offset_bcg}, the offsets of the BCG are plotted
against the radii of the innermost annulus. The offsets of the BCG are
smaller than the radius of the innermost annulus ($\sim30$ \% at the
maximum) except for 2A0335, suggesting that the difference in cluster
center between these two definitions did not affect our results
significantly.

\begin{figure}
 \begin{center}
  \FigureFile(70mm,50mm){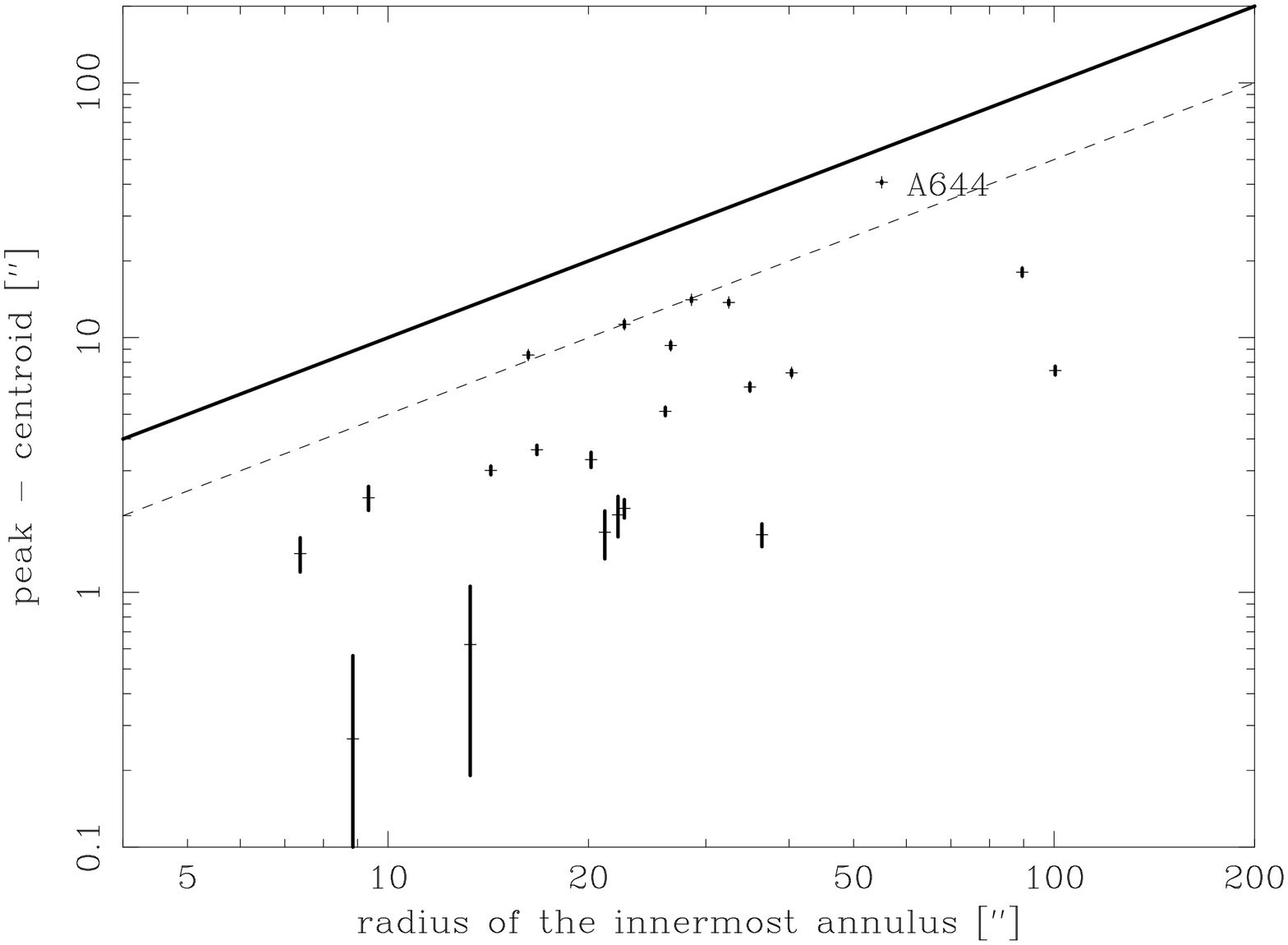} 
 \end{center}
\caption{Radius of the innermost annulus vs. offset between the X-ray
peak and the X-ray centroid. The dashed line represents 50 \% of the
radius of the innermost annulus.}
  \label{fig:offset_centro}
\end{figure}

We next examined the offset between the X-ray peak and the X-ray
centroid. To derive the X-ray centroid, we used the X-ray images in
which point sources were removed.  We replaced each embedded source with
the local diffuse X-ray emission surrounding the source by using the
CIAO task {\it dmfilth}.  This task replaces the counts within the
source regions with the values sampled from the background regions. As
the ACIS CCDs do not cover the entire X-ray emission for most clusters.
We thus define the X-ray centroid within a circular region that is
centered on the X-ray peak and has a radius that is tangent to the
detector edge. The offsets between the X-ray peak and the X-ray centroid
are plotted against the radii of the innermost annulus in
Figure \ref{fig:offset_centro}. The offsets of the X-ray centroid are slightly
larger than the offsets of the BCG but are smaller than the radii of the
innermost annulus. The offset is at most 74 \% (for A644) of the radius
of the innermost annulus, suggesting that the difference in these
definitions of a cluster center does not significantly affect the
results.

\subsection{Spherical Symmetry} 
\label{section:spherical_symmetry}

When we selected our sample, we excluded some clusters which are not
spherical symmetric in appearance. To quantify the spherical symmetry,
we measured the ellipticity ($\epsilon$) and the position angle ($PA$)
from the projected X-ray image. We used an iterative moment technique
derived from the treatment of the dispersion ellipse of the bivariate
normal frequency function of position vectors used by
\citet{carter:1980}. We first calculated the moments of the observed
X-ray images. From an image of $P$ pixels having $n_i$ counts in pixel
$i$, we computed the moment
\begin{equation}
 \mu_{mn} = \frac{1}{N}\sum_{i = 1}^{P} n_i (x_i - \bar{x})^m (y_i - \bar{y})^n ~~~ (m,n \leq 2),
\label{eqn:moments}
\end{equation}
where $N = \sum_{i = 1}^{P} n_i$, and ($\bar{x}$,$\bar{y}$) is the
centroid. Then ellipticity $\epsilon$ is 
\begin{equation}
 \epsilon = 1 - \frac{\Lambda_{-}}{\Lambda_{+}},
\label{eqn:ellipt}
\end{equation}
and the position angle of the major axis measured north through east in
celestial coordinates is 
\begin{equation}
 PA = \tan^{-1} (\frac{\mu_{11}}{\Lambda_{+}^2-\mu_{02}})+\frac{\pi}{2},
\label{eqn:PA}
\end{equation}
where $\Lambda_{\pm}$ ($\Lambda_{+} \geq \Lambda_{-}$) are the positive
roots of the quadratic equation
\begin{eqnarray}
\left|
\begin{array}{cc}
 \mu_{20} - \Lambda^2 & \mu_{11}  \\
 \mu_{11} & \mu_{02} - \Lambda^2  \\
\end{array}
\right|
= 0.
\label{eqn:lamda}
\end{eqnarray}

\begin{figure}
 \begin{center}
  \FigureFile(70mm,50mm){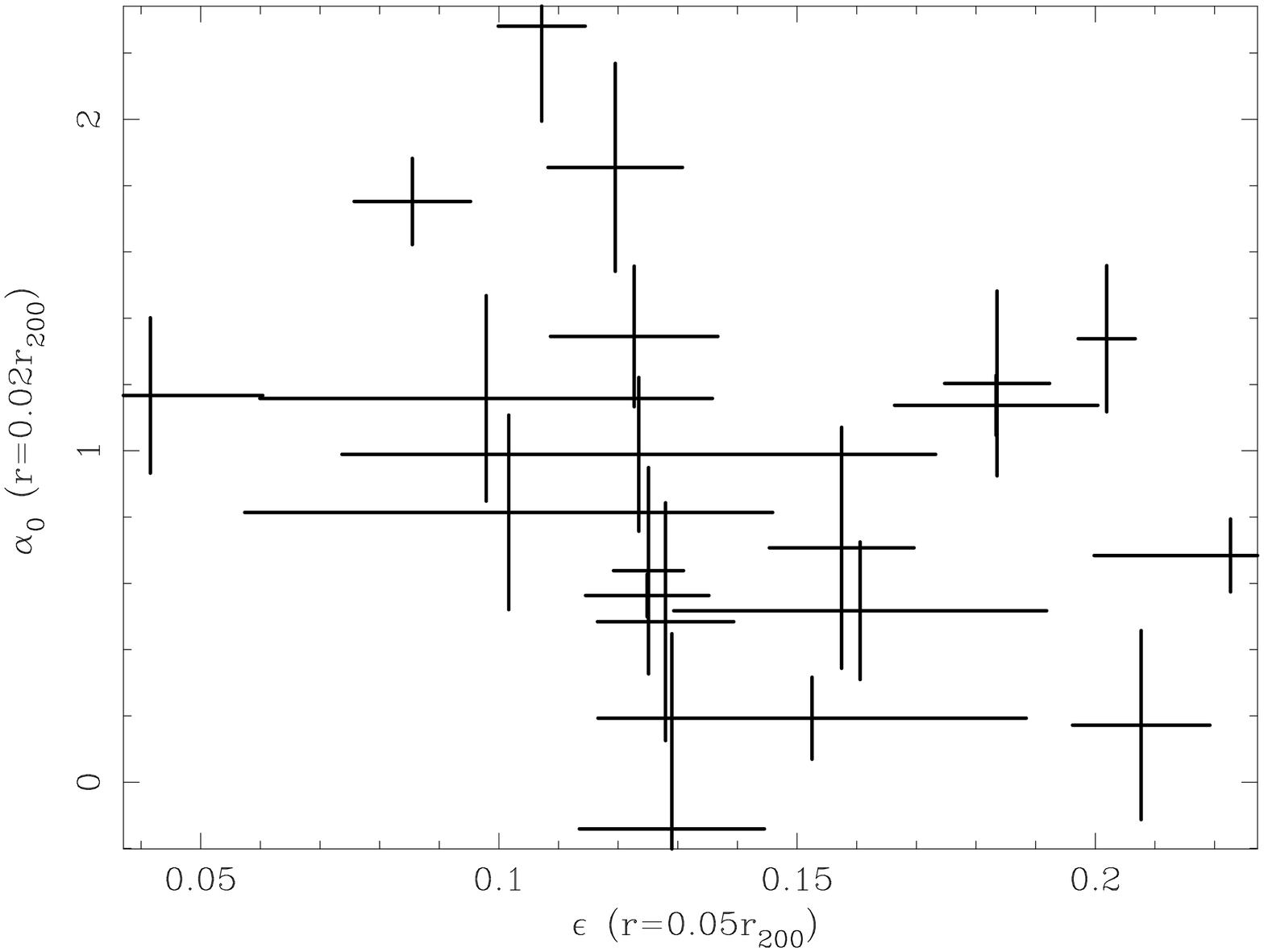} 
 \end{center}
  \caption{Ellipticity $\epsilon$ vs. inner slope $\alpha_0$. The errors are 1$\sigma$ (68.3 \%) confidence level. The correlation coefficient is $-0.31$ for this relation.}
\label{fig:ellipt_alpha}
\end{figure}

As in the case of the determination of the X-ray centroids, we used the
X-ray images from which point sources were removed and replaced with
local diffuse X-ray emission surrounding the source. We also employed
the \rosat\ images to determine the ellipticities and position angles in
outer regions of clusters. From the \rosat\ image, the point sources
were removed but the holes were not replaced with the background.  We
consider that if the ellipticity of a cluster affects the inner slope
$\alpha_0$, the ellipticity $\epsilon$ and the inner slope $\alpha_0$ will
show some kind of relation.  However, the plot in Figure
\ref{fig:ellipt_alpha} shows no correlation between the ellipticity
$\epsilon$ at $r=0.05r_{200}$ and the inner slope $\alpha_0$ (the
correlation coefficient is -0.31).  We therefore conclude that there was
no evidence that deviation from spherical symmetry affects the results
on the inner slope $\alpha_0$, although we cannot obtain no evidence
regarding symmetry along the line of sight.

\subsection{Central Structure}
\label{section:central_structure}

Recent \chandra\ observations have revealed remarkable structures in the
hot gas of the central region of some cooling flow clusters. These
structures may have observably affected the temperature and gas density
in our measurement. Furthermore, if these structures indicate a break of
hydrostatic equilibrium, they may systematically have affected the mass
profile we obtained. For 10 clusters in our sample, the presence of
central X-ray structures such as a cavity, hole, or plume has been
reported with \chandra\ observations in the literature (Table
\ref{tbl:central_structure}).

\begin{figure}
 \begin{center}
  \FigureFile(70mm,50mm){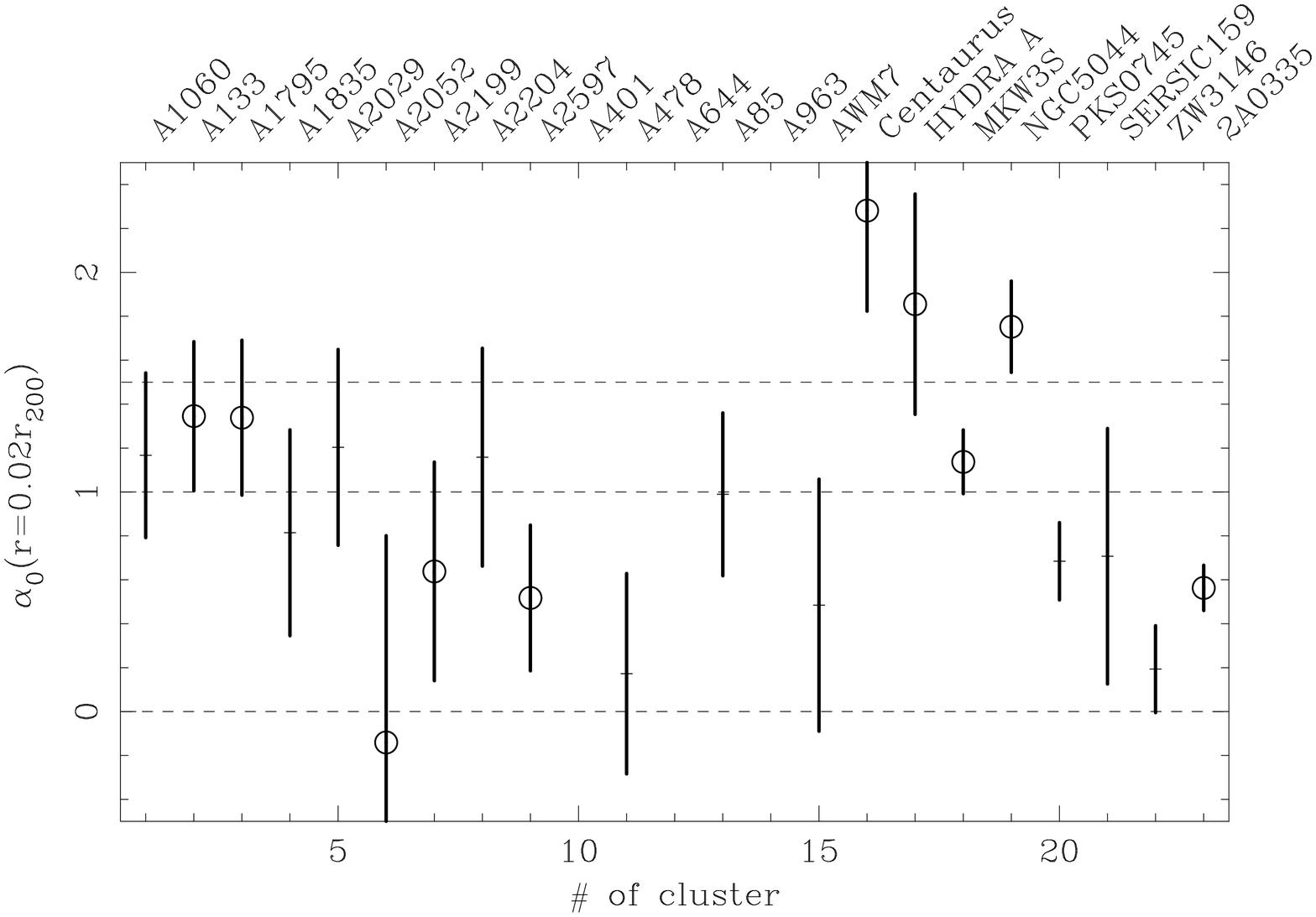} 
 \end{center}
  \caption{Inner slope $\alpha_0$ of mass
  profiles. The circles indicate 10 clusters in the literatures to
  have the central structure (see Table \ref{tbl:central_structure}).}
  \label{fig:ellipt_alpha_agn}
\end{figure}

In Figure \ref{fig:ellipt_alpha_agn}, we show the inner slope $\alpha_0$
again, indicating 10 clusters for which the central structure has been
found by open circles. It is found that the three clusters in which
$\alpha_0$ is as steep as 2 have central structures, and the range of
$\alpha_0$ becomes narrower if we neglect them. However, the
distribution of $\alpha_0$ from 0 to 1.2 is similar for clusters with
and those without central structures, though this is difficult to
conclude quantitatively.  Note that some of the central structures, such
as those in A133, A2597, and MKW3S are small enough to be removed from
the analysis. In order to evaluate the observational effect of these
structures, we removed the region of these structures from those
clusters and confirmed that the mass profiles were not significantly
affected.

\section{Relations between Inner Slope and Other Observational Parameters}
\label{section:relation_obs_param}

As shown in Section \ref{section:central_mass_slope}, the inner slope
$\alpha_0$ of the density profiles spans a wider range than that
estimated from their errors. Even if we neglect the three clusters in
which central structures might affect the results, $\alpha_0$ ranges
from 0 to 1.2 and spreads toward a flatter side than expected based on CDM
simulations. If this spread of the distribution is intrinsic, what is it
that determines the inner slope of the density profile?  In this
section, we explore observational parameters that primarily determine
the inner slope $\alpha_0$, by examining their correlations.

\subsection{Redshift vs. $\alpha_0$}

\begin{figure}
 \begin{center}
  \FigureFile(70mm,50mm){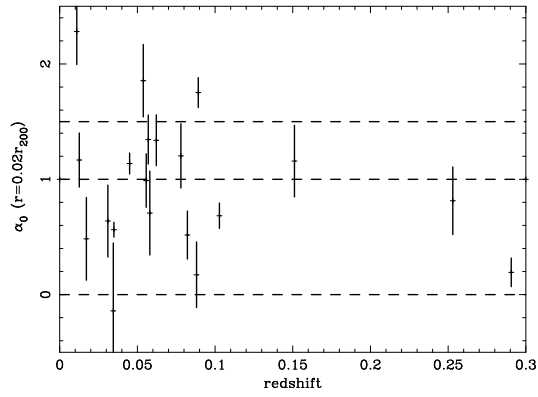} 
 \end{center}
  \caption{ Redshift vs. inner slope $\alpha_0$. The correlation
  coefficient is -0.25 for this relation.}  
  \label{fig:alpha0-redshift}
\end{figure}

We first show the relation between the redshift and the inner slope
$\alpha_0$ in Figure \ref{fig:alpha0-redshift}. The correlation
coefficient is -0.25 for this relation. Note that the range of
redshifts in our sample may not be sufficient to investigate the effect
of cosmological evolution.

\subsection{Temperature vs. $\alpha_0$}

\begin{figure}
 \begin{center}
 \FigureFile(70mm,50mm){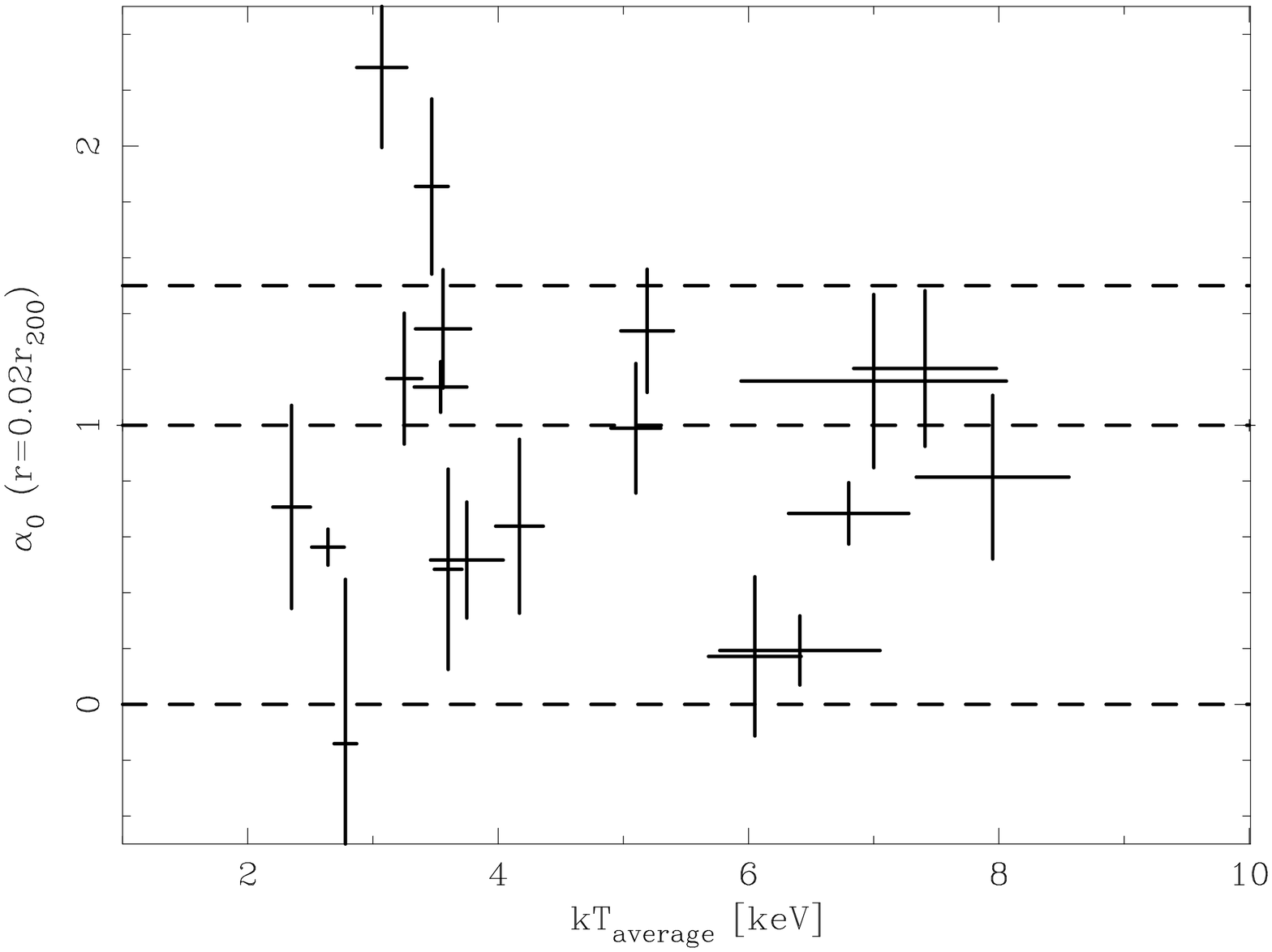} 
 \end{center}
  \caption{ Temperature vs. inner slope $\alpha_0$. The
correlation coefficient is 0.01 for this relation.}  
  \label{fig:alpha0-temp}
\end{figure}

We plot the spatially averaged temperature $kT_{\rm average}$ against
the inner slope $\alpha_0$ in Figure \ref{fig:alpha0-temp}. The
correlation coefficient is 0.01. Simple arguments based on virial
theorem suggest that the mass of a cluster is simply related to the
cluster temperature as $M \propto T^{3/2}$. This relation implies that
the inner slope $\alpha_0$ is not related to the scale of the cluster.

\subsection{Gas Fraction vs. $\alpha_0$}

We next investigated the relation between the gas fraction and the inner
slope $\alpha_0$. The gas fraction is the ratio of the hot gas mass to
the total mass, and is defined as a function of radius.  The integrated
gas mass profile $M_{\rm gas}(<r)$ is given by
\begin{eqnarray}
 \nonumber M_{\rm gas}(<r) & = & \int^{r}_{0} 4 \pi r^{\prime 2} \rho_{\rm gas}(r) dr^{\prime}\\
 & = &  4 \pi \mu m_{p} \int^{r}_{0} 4 \pi r^{\prime 2} n_{\rm gas}(r) dr^{\prime},
\end{eqnarray}
where $n_{\rm gas}(r)$ is the total number density of electrons and
ions, $\mu (= 0.6)$ is the mean molecular weight, and $m_p$ is the
proton mass. The gas fraction is defined as
\begin{equation}
 f_{\rm gas}(r) \equiv \frac{M_{\rm gas}(<r)}{M(<r)}.
\end{equation}

\begin{figure}
 \begin{center}
  \FigureFile(170mm,220mm){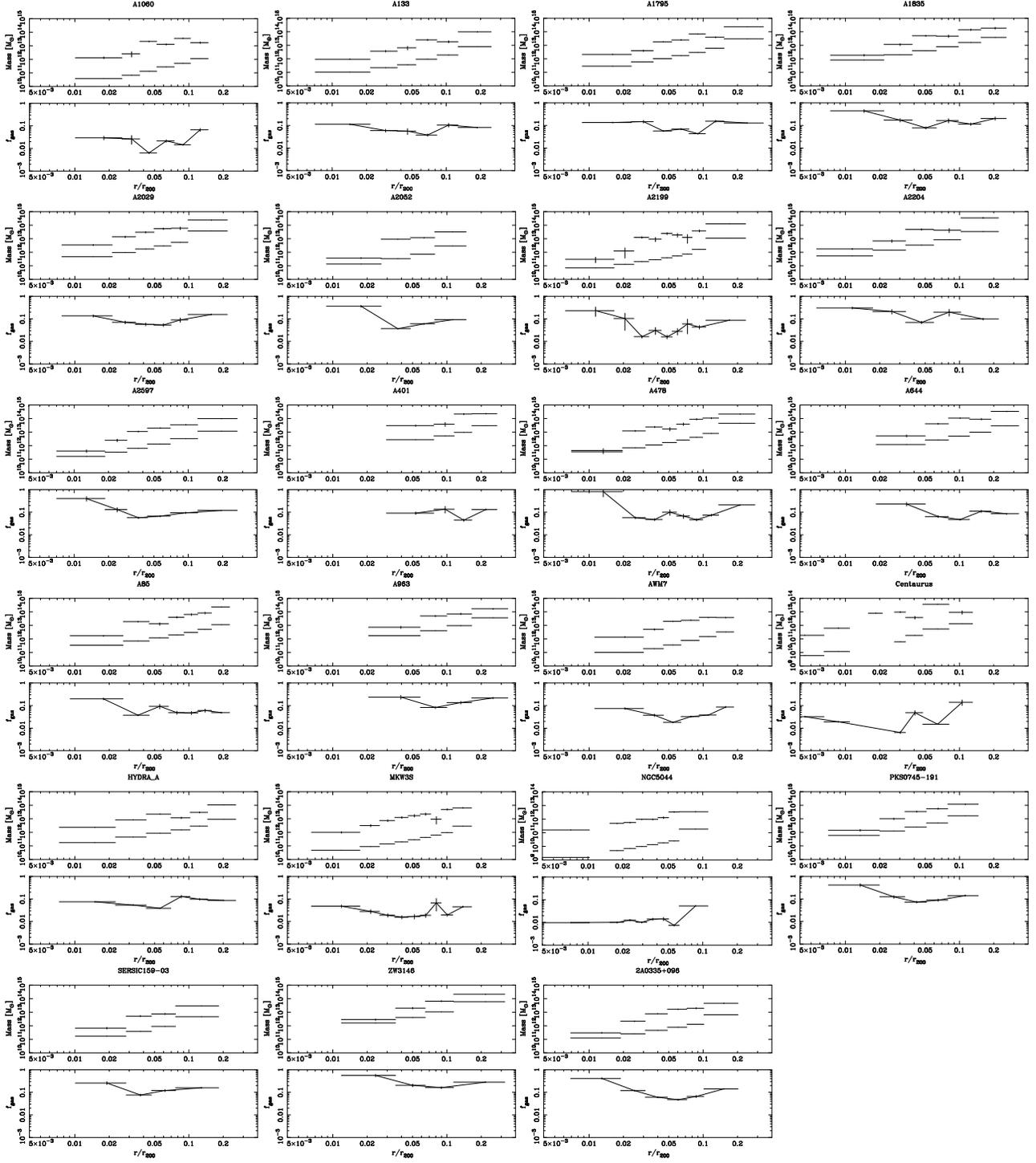}
 \end{center}
  \caption{Total integrated mass, gas mass, and gas fraction profiles of
  the 23 sample clusters. The radius is scaled with $r_{200}$. We show
  the total integrated mass and gas mass profiles in the upper panel,
  and the gas fraction profile in lower panel.}
  \label{fig:gas_mass_all}
\end{figure}

\begin{figure}
 \begin{center}
 \FigureFile(70mm,50mm){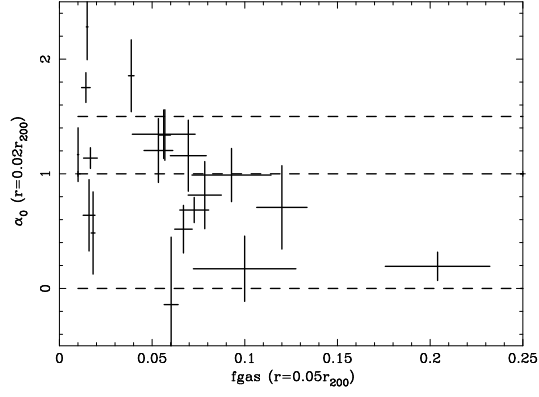} 
 \end{center}
  \caption{Gas fraction at the radius of $r = 0.05r_{200}$ vs. inner
  slope $\alpha_0$. The correlation coefficient is $-0.51$ for this
  relation. }  
\label{fig:alpha0-fgas}
\end{figure}

In Figure \ref{fig:gas_mass_all}, we present the profiles of total mass,
gas mass, and gas fraction for 23 sample clusters. It was found that the
gas fraction increases toward the center for some clusters. 
In Figure \ref{fig:alpha0-fgas}, the gas fractions at the radius of $r =
0.05r_{200}$ are plotted against the inner slope $\alpha_0$. A negative
correlation was observed with the correlation coefficient of $-0.51$,
for which case, at a significance level of about 3\%, the hypothesis of
no correlation is rejected. This correlation might be a kind of artifact
in the analysis, since we derive both the total mass profile and the gas
mass profile from the same gas density profile and gas temperature
profile.  However, it is unlikely that the observed correlation is due
to correlated errors between the two parameters, considering the size of
the errors.

\begin{figure}
 \begin{center}
 \FigureFile(70mm,50mm){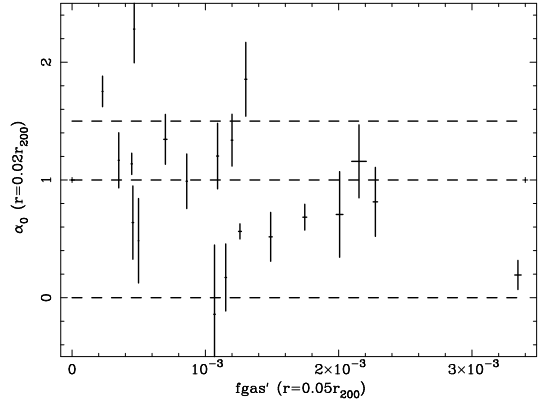} 
 \end{center}
  \caption{Gas fraction redefined in Equation \ref{eqn:fgas_dash} ($f_{\rm gas}^{\prime}$) vs. inner slope $\alpha_0$. The correlation coefficient is $-0.40$ for this relation.}  
\label{fig:alpha0-fgas_m200}
\end{figure}

In order to examine this correlation is artifact or not, we took
$M_{200}$, which is determined solely from the gas temperature and redshift,
instead of the integrated mass profile. We redefined the gas fraction
as
\begin{equation}
 f_{\rm gas}^{\prime}(r) \equiv \frac{M_{\rm gas}(<r)}{M_{200}}.
\label{eqn:fgas_dash}
\end{equation}
We plot $f_{\rm gas}^{\prime}$ at the radius of $r = 0.05r_{200}$
against the inner slope $\alpha_0$ in Figure \ref{fig:alpha0-fgas_m200}.
Although the correlation coefficient of -0.40 is smaller 
than that for $f_{\rm gas}$, the no correlation hypothesis is 
rejected at a significance level of less than 10\%.   

Correlations between the inner slope $\alpha_0$ and the gas fraction
according indicate that gas-rich clusters in the central region tend to
have a flat core: $\alpha<1$.

\section{Summary and Discussion}
\label{section:summary}

We have analyzed the \chandra\ data of 23 clusters of galaxies in order to
investigate central mass distribution. The high spatial resolution
imaging spectroscopy of \chandra\ and a new deprojection technique enable
us to measure the temperatures and gas densities in the very central
region of sample clusters without assuming any particular models. Under
the assumptions of hydrostatic equilibrium and spherical symmetry, we
obtained the deprojected mass profiles. 
Our major results are as follows.

\begin{enumerate}

 \item  
 The mass profiles scaled with $r_{200}$ and $M_{200}$ agree each
 other on the large scale $r>0.1r_{200}$. In contrast, the central
 ($r<0.1r_{200}$) mass profiles show a large scatter.

 \item 
 The inner slope $\alpha_0$ of the density profile was derived by
 fitting the mass profile with a general form of dark matter density
 profile for 20 clusters. The values of $\alpha_0$ span a wide range
 (from 0 to 2.3). For 6 out of 20 clusters, $\alpha_0$ are lower
 than unity at a 90 \% confidence level.

 \item 
 We investigated several features that might influence the results of
 the inner slope, including center position, ellipticity, and central
 structure of a cluster. We found that the systematic
 effects of these features are not significant except in the case of the
 central structure, which may broaden the distribution of the inner
 slope $\alpha_0$.  However, even if we excluded the clusters in which
 central structures were seen, the inner slope $\alpha_0$ distributes
 widely in the range from 0 to 1.2.

 \item 
 We examined the relationships between the inner slope $\alpha_0$ and other
 observational parameters. Although redshift, averaged temperature, and
 variation in the temperature profile are not correlated with the inner
 slope $\alpha_0$, gas fraction near the center of a cluster has a
 negative correlation with $\alpha_0$.

\end{enumerate}

CDM simulations predict that the inner slope
$\alpha$ is in the range $1<\alpha<2$. Therefore, our results are
inconsistent with the CDM simulations. Our observations provide flatter
slopes, at least for some clusters, than those expected from the CDM
simulations. This is true even if we neglect the clusters showing
central structures ($0 < \alpha_0 < 1.2$).

A similar claim has been presented by authors who are investigating the
rotation curve of galaxies and clusters. \citet{firmani:2001} examined
the observed rotation curves of dwarf and low surface brightness (LSB)
galaxies, and two clusters of galaxies, and found that all of those
objects have soft cores: $\alpha<1$. \citet{swaters:2003} observed the
rotation curve of 15 dwarf and LSB galaxies and found inner slopes in
the range of $0~^{<}_{\sim}~\alpha~^{<}_{\sim}~1$ in the majority.  This
inconsistency between observations and simulations in terms of the dark
matter distribution at the center of galaxies or clusters is called the
core problem.  Our results indicate that the core problem exists in a
significant fraction of clusters observed through X-ray
observations.

We also found a negative correlation between the inner slope and the gas
fraction. In the central region of clusters, the baryonic components
such as hot gas and stars in a galaxy are not negligible. These baryonic
components are usually considered to follow the gravitational potential,
which is predominantly determined by dark matter distribution.
However, our results imply that these baryonic components affect the
dark matter distribution in the central region of some clusters.

\citet{el-zant:2001} argued that the core problem can be resolved within
the framework of the standard CDM model by considering the dynamical
friction (DF) between dark matter and gas.  El-zant et al. assumed that
the gas is not initially smoothly distributed but is, rather,
concentrated in clumps. Such gas clumps move through smooth dark matter
particles, lose energy to the central dark halo, and heat it up. This
leads to the puffing up of the central regions and to the flattening of
the density profile. Monte Carlo simulation by El-zant et
al. successfully reproduced the observed flat density profile.  In the
DF model, the baryonic (gas) component is more centrally concentrated
than dark matter, because gas gives its energy to dark matter and
shrinks toward the center. So this model accounts for our observational
result that gas rich clusters in the central region tend to have a flat
core. 
Note that another numerical simulations including baryonic
components show the cooling baryons adiabatically compress the dark
matter and make a steeper central density profile
(e.g., \authorcite{blumenthal:1986} \yearcite{blumenthal:1986}). However,
this result is not consistent with our results because the gas rich
cluster in the central region is expected to have steeper profile if the
cooling baryons adiabatically compress the dark matter.

Although the DF model is one of many interpretations proposed to
account for the core problem, our results indicate that the numerical
simulation need to reconsider the treatment of the baryonic components
in galaxies or galaxy clusters.
\\
 

We thank all the staff members involved in the \chandra\ project. We
acknowledge Jone Arabadjis and Mark Bautz for their support on the
deprojection technique and their useful comments.

\clearpage

\begin{table}
  \caption{Observation log of the sample clusters.}  \label{tbl:obs_log}
\begin{center}
 \begin{tabular}{lllllllll}\hline
  ID & Cluster & Obs. & Ra (deg)$^a$ & Dec (deg)$^a$ & Observation date &
  Exp.  & Screened$^b$  & ACIS$^c$ \\
  & & ID & & & & (ks) & Exp. (ks) & chip \\ \hline\hline
 1 & A1060 & 2220 & 159.073 & -27.569 & 2001-06-04 04:43:23 & 31.9 & 30.0& I \\
 2 & A133 & 2203 & 15.689 & -21.882 & 2000-10-13 22:27:02 & 35.5 & 34.5& S \\
 3 & A1795 & 493 & 207.205 & 26.608 & 2000-03-21 07:54:49 & 19.6 & 19.6& S \\
 & 	 & 3666 & 207.204 & 26.576 & 2002-06-10 16:21:19 & 14.4 & 13.8& S \\
 & 	 & 494 & 207.236 & 26.607 & 1999-12-20 05:00:57 & 19.5 & 17.6& S \\
 4 & A1835 & 495 & 210.272 & 2.895 & 1999-12-11 16:48:33 & 19.5 & 19.5& S \\
 & 	 & 496 & 210.222 & 2.867 & 2000-04-29 06:55:44 & 10.7 & 10.7 & S \\
 5 & A2029 & 891 & 227.725 & 5.764 & 2000-04-12 06:38:56 & 19.8 & 19.8& S \\
 6 & A2052 & 890 & 229.182 & 7.012 & 2000-09-03 06:01:22 & 36.8 & 36.8& S \\
 7 & A2199 & 498 & 247.188 & 39.553 & 1999-12-11 10:47:37 & 18.9 & 18.9& S \\
 & 	 & 497 & 247.135 & 39.560 & 2000-05-13 17:36:15 & 19.5 & 17.9& S \\
 8 & A2204 & 499 & 248.185 & 5.557 & 2000-07-29 02:49:42 & 10.1 & 10.1& S \\
 9 & A2597 & 922 & 351.337 & -12.135 & 2000-07-28 05:13:47 & 39.4 & 25.1& S \\
 10 & A401 & 518 & 44.727 & 13.579 & 1999-09-17 21:35:26 & 18.0 & 18.0& I \\
 & 	& 2309 & 44.732 & 13.461 & 2000-11-03 19:10:36 & 11.6 & 11.6& I \\
 11 & A478 & 1669 & 63.362 & 10.436 & 2001-01-27 03:28:03 & 42.4 & 42.4& I \\
 12 & A644 & 2211 & 124.329 & -7.543 & 2001-03-26 00:27:49 & 29.7 & 29.2& S \\
 13 & A85 & 904 & 10.442 & -9.374 & 2000-08-19 07:06:52 & 38.4 & 38.4& I \\
 14 & A963 & 903 & 154.284 & 39.063 & 2000-10-11 00:01:18 & 36.3 & 36.3& S \\
 15 & AWM7 & 908 & 43.665 & 41.664 & 2000-08-19 18:30:01 & 47.9 & 47.9& I \\
 16 & Centaurus & 504 & 192.207 & -41.334 & 2000-05-22 00:33:17 & 31.7 & 28.8& S \\
 & 	     & 505 & 192.199 & -41.334 & 2000-06-08 00:51:50 & 10.0 & 10.0& S \\
 17 & Hydra A & 575 & 139.527 & -12.091 & 1999-10-30 07:29:02 & 23.8 & 23.8& I \\
 & 	   & 576 & 139.527 & -12.091 & 1999-11-02 11:31:54 & 19.5 & 19.5& S \\
 18 & MKW3S & 900 & 230.488 & 7.757 & 2000-04-03 12:26:13 & 57.3 & 57.3& I \\
 19 & NGC5044 & 798 & 198.859 & -16.378 & 2000-03-19 15:42:42 & 20.5 & 19.8& S \\
 20 & PKS0745-191 & 2427 & 116.860 & -19.306 & 2001-06-16 05:32:52 & 17.9 & 17.9& S \\
 & 	       & 508 & 116.870 & -19.277 & 2000-08-28 22:15:31 & 28.0 & 4.6& S \\
 21 & Sersic159-03 & 1668 & 348.515 & -42.713 & 2001-08-13 16:41:20 & 9.9 & 9.9& S \\
 22 & ZW3146 & 909 & 155.905 & 4.166 & 2000-05-10 03:20:25 & 46.0 & 46.0& I \\
 23 & 2A0335+096 & 919 & 54.666 & 10.008 & 2000-09-06 00:03:13 & 19.7 & 14.1&
  S \\ \hline
 \end{tabular}
\end{center}
 {\footnotesize 
 \noindent
 $^{a}$ Nominal pointing position of the observation in
 Equinox 2000.0 \\
 $^{b}$ Exposure time after lightcurve screening (see $\S$4.2) \\
 $^{c}$ Detector on the aim point \\} 
\end{table}

\clearpage

\begin{table}
  \caption{Properties of the sample clusters. We show the redshifts,
  hydrogen column densities of the galactic absorption, temperatures,
  and X-ray fluxes of the sample clusters.  The temperatures are
  referred to \citet{reiprich:2002}, \citet{ota:2000}, and
  \citet{allen:1996}. } 
 \label{tbl:sample_properties} 
\begin{center}
 \begin{tabular}{llllll}\hline
 Cluster & redshift & $N_{\rm H}^a$ & kT & $f_{\rm X}$$^b$  & Ref. $^c$  \\
 & & [$10^{20}$ cm$^{-2}$] & [keV] & [$10^{-11}$ ergs s$^{-1}$ cm$^{-2}$] & \\ \hline\hline
 A1060 	     & 0.0126 & 4.79 & 3.24$^{+0.06}_{-0.06}$ & 9.95 & R\\
 A133  	     & 0.0570 & 1.55 & 3.80$^{+2.00}_{-0.90}$ & 2.12 & R\\
 A1795 	     & 0.0622 & 1.20 & 7.80$^{+1.00}_{-1.00}$ & 6.27 & R\\
 A1835 	     & 0.2530 & 2.30 & 7.42$^{+0.61}_{-0.43}$ & 1.47 & O\\
 A2029 	     & 0.0780 & 3.07 & 9.10$^{+1.00}_{-1.00}$ & 6.94 & R\\
 A2052 	     & 0.0345 & 2.78 & 3.03$^{+0.04}_{-0.04}$ & 4.71 & R\\
 A2199 	     & 0.0310 & 0.87 & 4.10$^{+0.08}_{-0.08}$ & 10.64 & R\\
 A2204 	     & 0.1511 & 5.66 & 7.21$^{+0.25}_{-0.25}$ & 2.75 & R\\
 A2597 	     & 0.0822 & 2.50 & 4.40$^{+0.40}_{-0.70}$ & 2.21 & R\\
 A401  	     & 0.0748 & 10.3 & 8.00$^{+0.40}_{-0.40}$ & 5.28 & R\\
 A478  	     & 0.0881 & 14.8 & 8.40$^{+0.80}_{-1.40}$ & 5.15 & R\\
 A644  	     & 0.0704 & 6.95 & 7.90$^{+0.80}_{-0.80}$ & 4.02 & R\\
 A85   	     & 0.0557 & 3.37 & 6.90$^{+0.40}_{-0.40}$ & 7.43 & R\\
 A963  	     & 0.2057 & 1.40 & 6.83$^{+0.51}_{-0.51}$ & 0.59 & O\\
 AWM7  	     & 0.0172 & 9.91 & 3.75$^{+0.09}_{-0.09}$ & 1.58 & R\\
 Centraurus   & 0.0110 & 8.07 & 3.68$^{+0.06}_{-0.06}$ & 27.19 & R\\
 Hydra A      & 0.0538 & 4.90 & 4.30$^{+0.40}_{-0.40}$ & 4.78 & R\\
 MKW3S        & 0.0450 & 3.04 & 3.70$^{+0.20}_{-0.20}$ & 3.30 & R\\
 NGC5044      & 0.0089 & 5.03 & 1.07$^{+0.01}_{-0.01}$ & 5.51 & R\\
 PKS0745-191  & 0.1028 & 40.7 & 7.21$^{+0.11}_{-0.11}$ & 2.44 & R\\
 Sersic159-03 & 0.0580 & 1.76 & 3.00$^{+1.20}_{-0.70}$ & 2.49 & R\\
 ZW3146       & 0.2906 & 2.94 & 6.10$^{+0.30}_{-0.30}$ & 0.66 & A\\
 2A0335+096   & 0.0349 & 17.6 & 3.01$^{+0.07}_{-0.07}$ & 9.16 & R\\ \hline
 \end{tabular} 
\end{center}
{\footnotesize 
\noindent
$^a$ Hydrogen column density of the galactic absorption. \\
$^b$ X-ray flux in units of $10^{-11}$ ergs s$^{-1}$ cm$^{-2}$. The
energy bands are 0.1-2.4 keV for R and 2-10 keV for O and A. \\
$^c$ References R:\citet{reiprich:2002}, O:\citet{ota:2000}, and
A:\citet{allen:1996}.
}
\end{table}

\clearpage

\begin{table}
\caption{Fitting results of the
 temperature profiles with the exponential + constant model. } 
 \label{table:fit_temperature_exp} 
\begin{center}
 \begin{tabular}{llllll}\hline\hline
  Cluster & $T_0$ [keV]$^a$ & $T_1$ [keV]$^a$ & $r_T^a$ [kpc]& $\chi^2_{\rm stat}$/dof$^b$   
  & Systematic \\ 
  & & & & & error$^c$ \\ \hline
A1060 & 3.16 $\pm$ 0.22 & -0.00 $\pm$ 0.80  & 850.0 $\pm$ 918.5      & 126.4/4& 0.183 \\
A133 & 4.52 $\pm$ 0.36 & -3.45 $\pm$ 0.33   & 90.9 $\pm$ 28.5        & 20.7/4 & 0.082 \\
A1795 & 6.03 $\pm$ 0.51 & -3.36 $\pm$ 0.57  & 151.4 $\pm$ 63.1       & 105.1/5& 0.099 \\
A1835 & 10.10 $\pm$ 0.69 & -7.60 $\pm$ 0.56 & 103.4 $\pm$ 28.2       & 10.2/4 & .. \\
A2029 & 9.89 $\pm$ 0.30 & -5.73 $\pm$ 0.34  & 144.4 $\pm$ 22.7      &
  3.9/4 & .. \\
A2052 & 3.20 $\pm$ 0.27 & -5.67 $\pm$ 8.39  & 14.9 $\pm$ 22.4       & 24.8/2 & 0.734 \\
A2199 & 4.92 $\pm$ 0.32 & -3.34 $\pm$ 0.41  & 57.2 $\pm$ 22.2        & 97.3/7 & 0.093 \\
A2204 & 8.45 $\pm$ 1.29 & -7.21 $\pm$ 1.80  & 58.1 $\pm$ 34.0        & 26.1/3 & 0.235 \\
A2597 & 6.02 $\pm$ 0.39 & -4.33 $\pm$ 0.35  & 162.7 $\pm$ 29.1       &  6.6/4 & .. \\
A401 & 7.71 $\pm$ 0.62 & 0.00 $\pm$ 29.09   & 12.9 $\pm$ 0.1         & 19.0/2 & 0.170 \\
A478 & 7.03 $\pm$ 0.35 & -5.52 $\pm$ 0.78   & 48.4 $\pm$ 20.0        & 33.6/6 & 0.076 \\
A644 & 6.47 $\pm$ 0.39 & -28.72 $\pm$ 28.72 & 14.9 $\pm$ 27.3        & 27.6/3 & 0.128 \\
A85 & 6.44 $\pm$ 0.27 & -4.56 $\pm$ 0.46    & 78.4 $\pm$ 21.6        & 26.0/5 & 0.065 \\
A963 & 6.22 $\pm$ 0.25 & -29.78 $\pm$ 23.80 & 13.3 $\pm$ 6.1         &  1.5/2 & .. \\
AWM7 & 3.78 $\pm$ 0.10 & -29.60 $\pm$ 27.94 & 7.2 $\pm$ 3.8          & 52.8/4 & 0.064 \\
Centaurus & 4.63 $\pm$ 0.90 & -3.92 $\pm$ 0.86 & 78.0 $\pm$ 35.2     & 1167.4/7& 0.215 \\
Hydra A & 3.80 $\pm$ 0.24 & -0.95 $\pm$ 0.28 & 113.3 $\pm$ 95.0      & 33.4/4 & 0.064 \\
MKW3S & 3.71 $\pm$ 0.14 & -1.10 $\pm$ 0.59 & 30.9 $\pm$ 23.2         & 42.3/7 & 0.078 \\
NGC5044 & 1.65 $\pm$ 0.39 & -1.10 $\pm$ 0.37 & 102.9 $\pm$ 54.5      & 168.6/7& 0.065 \\
PKS0745-191 & 11.16 $\pm$ 0.65 & -8.89 $\pm$ 0.57 & 154.6 $\pm$ 25.7 & 9.9/3  & .. \\
Sersic159-03 & 2.69 $\pm$ 0.09 & -1.31 $\pm$ 0.13 & 58.0 $\pm$ 15.7  & 1.8/2  & .. \\
ZW3146 & 8.04 $\pm$ 0.31 & -6.86 $\pm$ 0.58 & 57.2 $\pm$ 13.0        & 0.6/2  & .. \\
2A0335+096 & 3.96 $\pm$ 0.40 & -3.18 $\pm$ 0.35 & 97.1 $\pm$ 30.2    & 60.6/4 &
  0.089 \\ \hline
\end{tabular} 
\end{center}
{\footnotesize 
\noindent
$^{a}$ Errors are estimated by including a systematic error (see text). \\
$^{b}$ Original reduced $\chi^2$ without a systematic error
 ($\chi^2_{\rm stat}$).  \\
$^{c}$ Systematic error adopted so as to get the reduced $\chi^2$ value
 of unity. This error includes the fluctuations in the intrinsic temperature
profiles or unknown systematic errors in our analysis procedure. \\} 
\end{table}

\clearpage

\begin{table}
\caption{Fitting results of the gas density profiles with the NFW gas density model.}
\label{table:fit_gas_density_nfw} 
\begin{center}
\begin{tabular}{llllll}  \hline\hline
 Cluster        & $n_{\rm e0}$ [$10^{-2}$ cm$^{-3}$]$^a$ & $r_{\rm s}$ [kpc]$^a$ &
 $B^a$ & $\chi^2_{\rm stat}$/dof$^b$ & Systematic \\
 & & & &  & error$^c$ \\ \hline
 A1060 		& 0.83$\pm$0.09 	& 145.1$\pm$48.2 	& 5.13$\pm$1.08 	& 565.3/4    & 0.078 \\
 A133 		& 4.21$\pm$0.80 	& 91.6$\pm$21.9 	& 6.40$\pm$0.37 	& 1584.3/4   & 0.121 \\
 A1795 		& 4.76$\pm$0.45 	& 252.8$\pm$37.6 	& 8.54$\pm$0.50 	& 6460.0/5   & 0.084 \\
 A1835 		& 22.10$\pm$1.52 	& 112.1$\pm$9.3 	& 7.04$\pm$0.16 	& 257.3/4    & 0.051 \\
 A2029 		& 5.84$\pm$0.08 	& 193.7$\pm$4.5 	& 7.11$\pm$0.06 	& 31.9/5     & 0.011 \\
 A2052 		& 4.74$\pm$0.24 	& 64.5$\pm$5.3 	        & 5.72$\pm$0.13 	& 144.1/2    & 0.029 \\
 A2199 		& 3.00$\pm$0.26 	& 235.1$\pm$40.6 	& 8.52$\pm$0.66 	& 5112.0/7   & 0.079 \\
 A2204 		& 21.60$\pm$3.10 	& 89.3$\pm$14.4 	& 7.15$\pm$0.28 	& 557.3/3    & 0.10 \\
 A2597 		& 8.11$\pm$0.62 	& 157.4$\pm$17.6 	& 8.94$\pm$0.38 	& 537.9/4    & 0.071 \\
 A401 		& 0.71$\pm$0.03 	& 2842.0$\pm$664.9 	& 19.05$\pm$10.64 	& 97.6/2     & 0.035 \\
 A478 		& 6.46$\pm$1.25 	& 338.3$\pm$118.0 	& 9.99$\pm$1.53 	& 13901.8/6  & 0.155 \\
 A644 		& 1.63$\pm$0.17 	& 493.6$\pm$125.8 	& 8.30$\pm$1.05 	& 972.4/3    & 0.085 \\
 A85 		& 3.86$\pm$0.48 	& 122.5$\pm$21.6 	& 5.76$\pm$0.26 	& 2246.4/5   & 0.081 \\
 A963 		& 2.76$\pm$0.13 	& 477.2$\pm$58.5 	& 9.07$\pm$0.57 	& 39.3/2     & 0.033 \\
 AWM7 		& 1.26$\pm$0.14 	& 140.7$\pm$17.7 	& 5.00$\pm$0.26 	& 2899.1/4   & 0.083 \\
 Centaurus 	& 7.79$\pm$1.75 	& 24.7$\pm$5.5  	& 5.72$\pm$0.23 	& 6499.1/7   & 0.143 \\
 Hydra A 	& 7.03$\pm$1.53 	& 87.9$\pm$22.7 	& 6.67$\pm$0.43 	& 6555.0/4   & 0.151 \\
 MKW3S          & 3.23$\pm$0.33 	& 106.7$\pm$20.5 	& 5.95$\pm$0.41 	& 608.8/7    & 0.083 \\
 NGC5044 	& 3.98$\pm$0.95 	& 34.8$\pm$15.5 	& 6.22$\pm$0.95 	& 3186.8/7   & 0.163 \\
 PKS0745-191 	& 12.00$\pm$0.38 	& 145.6$\pm$8.1 	& 7.39$\pm$0.16 	& 72.3/3     & 0.022 \\
 Sersic159-03 	& 5.14$\pm$0.40 	& 240.5$\pm$54.8 	& 11.08$\pm$1.46 	& 101.5/2    & 0.057 \\
 ZW3146 	& 14.30$\pm$1.99 	& 258.2$\pm$63.0 	& 9.74$\pm$1.00 	& 473.6/2    & 0.096 \\
 2A0335+096 	& 9.23$\pm$1.30 	& 90.5$\pm$17.3 	& 7.41$\pm$0.47 	& 1887.5/4   & 0.111 \\  \hline
 \end{tabular}
\end{center}
{\footnotesize 
\noindent
$^{a}$ Errors are estimated by including a systematic error (see text). \\
$^{b}$ Reduced $\chi^2$ without a systematic error ($\chi^2_{\rm stat}$).  \\
$^{c}$ Systematic error adopted so as to make the reduced $\chi^2$ value
 unity.\\}
\end{table}

\clearpage

\begin{table}
\caption{Fitting results of the mass profile with the general form of density profile given by Equation \ref{eqn:general_dens_profile}. The errors are 90 \% confidence
  intervals.}
 \label{tbl:alpha_002r200}
\begin{center}
\begin{tabular}{lllll}\hline\hline
Cluster &  $r_s$ [kpc]$^a$ &
 $\alpha_0(0.02r_{200})$ $^a$ &  $\chi^2_{\rm stat}$/dof$^b$ & Systematic
 \\ 
 & & & & error$^c$ \\ \hline
A1060 &  6.0$\pm$4.2 &  1.17$\pm$0.38 &  120.3/3 & 0.224  \\
A133 & 1000 (fixed) &  1.32$\pm$0.38 & 31.8/3 & 0.398  \\
A1795 & 1000 (fixed) &  1.29$\pm$0.43 & 194.2/4 & 0.482  \\
A1835 & 51.1$\pm$200.1 &  0.82$\pm$0.47 &  13.9/3  &0.278  \\
A2029 & 760.0$\pm$1733.0 &  1.15$\pm$0.47 & 7.0/3 & ...  \\
A2052 & 3.4$\pm$8.1 &  -0.14$\pm$0.96 & 176.4/1 & 0.524  \\
A2199 & 163.3$\pm$8.2 &  0.64$\pm$0.46 &  36.8/6 & 0.440  \\
A2204 & 1000 (fixed) &  1.13$\pm$0.56 & 19.0/2 &0.525  \\
A2597 & 21.8$\pm$76.6 &  0.52$\pm$0.33 & 20.9/3 & 0.192  \\
A478 & 3.1$\pm$3.5 &  0.17$\pm$0.42 &  23.8/4 & 0.308  \\
A85 & 1000 (fixed) &  0.94$\pm$0.37 &  95.6/4 &0.365  \\
AWM7 & 7.1$\pm$8.1 &  0.48$\pm$0.39 & 70.7/3 & 0.290  \\
Centaurus & 10.0$\pm$20.8 &  2.28$\pm$0.46 & 549.7/4 & 0.640  \\
Hydra A & 1000 (fixed) &  1.85$\pm$0.55 & 237.3/3  &0.556  \\
MKW3S & 86.4$\pm$101.8 &  1.22$\pm$0.18 & 17.1/4 &  ... \\
NGC5044 & 1000 (fixed) &  1.76$\pm$0.21 & 32.8/5 &  0.217  \\
PKS0745-191 & 17.4$\pm$6.6 &  0.68$\pm$0.18 & 2.3/2 & ... \\
Sersic159-03 & 64.5$\pm$25.0 &  0.68$\pm$0.67 &  15.1/1 & 0.378  \\
ZW3146 & 69.4$\pm$9.0 &  0.19$\pm$0.20 &  0.3/1 & ... \\
2A0335+096 & 15.4$\pm$35.3 &  0.56$\pm$0.10 &
 11.9/3 & ... \\ \hline
\end{tabular}
\end{center}
{\footnotesize 
\noindent
$^{a}$ Errors are estimated by including a systematic error (see text). \\
$^{b}$ Reduced $\chi^2$ before including a systematic error ($\chi^2_{\rm stat}$).  \\
$^{c}$ Systematic error adopted so as to make the reduced $\chi^2$ value
 unity.\\} 
\end{table}

\clearpage

\begin{table}
\caption{Remarkable structures in the central region
 of the sample clusters in literatures. } 
\label{tbl:central_structure}
\begin{center}
\begin{tabular}{lll}\hline\hline
 Cluster    & Structure  & Reference  \\ \hline
 A133       & tongue     & \citet{fujita:2002}  \\
 A1795      & filament   & \citet{fabian:2001}  \\
 A2052      & holes      & \citet{blanton:2001} \\
 A2199      & depression & \citet{johnstone:2002} \\
 A2597      & cavities   & \citet{mcnamara:2001} \\
 Centaurus & plume      & \citet{sanders:2002} \\
 Hydra A    & depression & \citet{mcnamara:2000} \\
 NGC5044    & hole       & \citet{buote:2003} \\
 MKW3S      & filament \& depression & \citet{mazzotta:2002} \\
 2A0335+096 & cavities \& blobs  & \citet{mazzotta:2003}\\  \hline
\end{tabular}
\end{center}
\end{table}

\clearpage

\clearpage

\clearpage

\clearpage

\clearpage

\end{document}